\documentclass[prd,twocolumn,preprintnumbers,superscriptaddress,nofootinbib]{revtex4}

\usepackage{amstext,amssymb}
\usepackage{amsmath}
\usepackage{graphicx}
\usepackage[hyperfootnotes=true]{hyperref}
\usepackage{xspace}
\usepackage{color}
\usepackage{slashed} 
\usepackage{multirow}
\usepackage[normalem]{ulem}

\begin{document}
\preprint{IP/BBSR/2018-7}
\title{Baryogenesis via Leptogenesis from Asymmetric Dark Matter and radiatively generated Neutrino mass}
\author{Nimmala Narendra}
\email{ph14resch01002@iith.ac.in}
\affiliation{Indian Institute of Technology Hyderabad, Kandi, Sangareddy, 502285, Telangana, India}
\author{Sudhanwa Patra}
\email{sudha.astro@gmail.com}
\affiliation{Indian Institute of Technology Bhilai, Raipur, Chhatishgarh, India}
\author{Narendra Sahu}
\email{nsahu@iith.ac.in}
\affiliation{Indian Institute of Technology Hyderabad, Kandi, Sangareddy, 502285, Telangana, India}
\author{Sujay Shil}
\email{sujay@iopb.res.in}
\affiliation{Institute of Physics, Sachivalaya Marg, Bhubaneswar, Odisha 751005, India}
\affiliation{Homi Bhabha National Institute, Training School Complex, Anushakti Nagar, Mumbai 400085, India}
\begin{abstract}
We propose an extension of the standard model (SM) by including a dark sector comprising of three generations 
of heavy right-handed neutrinos, a singlet scalar and a singlet Dirac fermion, where the latter two particles are 
stable and are viable candidates of dark matter (DM). In the early Universe, the CP-violating out-of-equilibrium 
decay of heavy right-handed neutrinos to singlet Dirac fermion and scalar in the dark sector generates a net 
DM asymmetry. The latter is then transported to the visible sector via a dimension eight operator which conserves 
$B-L$ symmetry and is in thermal equilibrium above the sphaleron decoupling temperature. An additional light singlet 
scalar is introduced which mixes with the SM Higgs and pave a path for annihilating the symmetric components of the 
DM candidates. Then we discuss the constraints on singlet-doublet Higgs mixing from invisible Higgs 
decay, signal strength at LHC and direct search of DM at terrestrial laboratories. At tree level the neutrinos are 
shown to be massless since the symmetry of dark sector forbids the interaction of right-handed neutrinos with the SM 
particles. However, at one loop level the neutrinos acquire sub-eV masses as required by the oscillation experiments.
\end{abstract}
\maketitle
\section{Introduction} \label{Intro}
The evidence from galaxy rotation curve, gravitational lensing and large scale structure of the Universe irrefutably 
proven the existence of dark matter (DM) in a large scale ($\gtrsim$ a few kpc)~\cite{dm_review}. However, the microscopic picture 
of DM is hitherto not known. The only piece of information that we know about the DM is its relic abundance and is precisely 
measured by the satellite borne experiments WMAP~\cite{Hinshaw:2012aka} and PLANCK~\cite{Ade:2015xua} to be $\Omega_{\rm DM}h^2 
= 0.1199\pm 0.0027$. However, a little is known about the underlying mechanism of generating relic abundance of DM. The 
most considered scenario is the DM to be a Weakly Interacting Massive Particle (WIMP)~\cite{kolbturner}. The latter gets 
thermalised in the early Universe due to its weak interaction property. As the temperature falls below its mass scale, the 
DM gets decoupled from the thermal bath and its density in a comoving volume remain constant which we measure today. This 
is usually referred as WIMP miracle. 

A curious observation about DM is that its relic density is about 5 times larger than the baryon density of the present Universe, 
{\it i.e.} $\Omega_{\rm DM} \approx 5 \Omega_B$. This implies that the relic density of DM can be generated in a similar way that 
baryon asymmetry of the Universe has been generated. See for example~\cite{old_DM_Baryon_asy,Asydm_models1,Asydm_models2,Asydm_models3,
Asydm_review}. The observed baryon asymmetry, usually reported in terms of the baryon to photon ratio, $\eta=n_{B}/n_{\gamma}$, is 
given as \cite{Patrignani:2016xqp},
\begin{equation}
5.8 \times 10^{-10} \leq \eta \leq 6.6 \times 10^{-10} \hspace{0.3cm} (BBN)\hspace{0.4cm} (95 \% CL)\,,
\label{Y_{B}}
\end{equation}
where $\eta = 7.04 Y_B$ with $Y_{B} \equiv n_{B}/s$. Similarly the observed DM abundance can be expressed as 
\begin{equation}
Y_{\rm DM}\equiv \frac{n_{\rm DM}}{s}= 4 \times 10^{-10} \left( \frac{1 \rm GeV}{M_{DM}}\right)\left(\frac{\Omega_{\rm DM}h^2}{0.11}  \right)\,.
\end{equation} 
This implies that $Y_{\rm DM}/Y_B \approx {\cal O}(1)$ if $M_{\rm DM} \sim 5 {\rm GeV}$. However, it can vary from a GeV to TeV depending 
on the magnitude of CP violation in visible and dark sectors. See for instance~\cite{Asydm_models3}.

\begin{figure}[h!]
				\centering
				\includegraphics[width = 80mm]{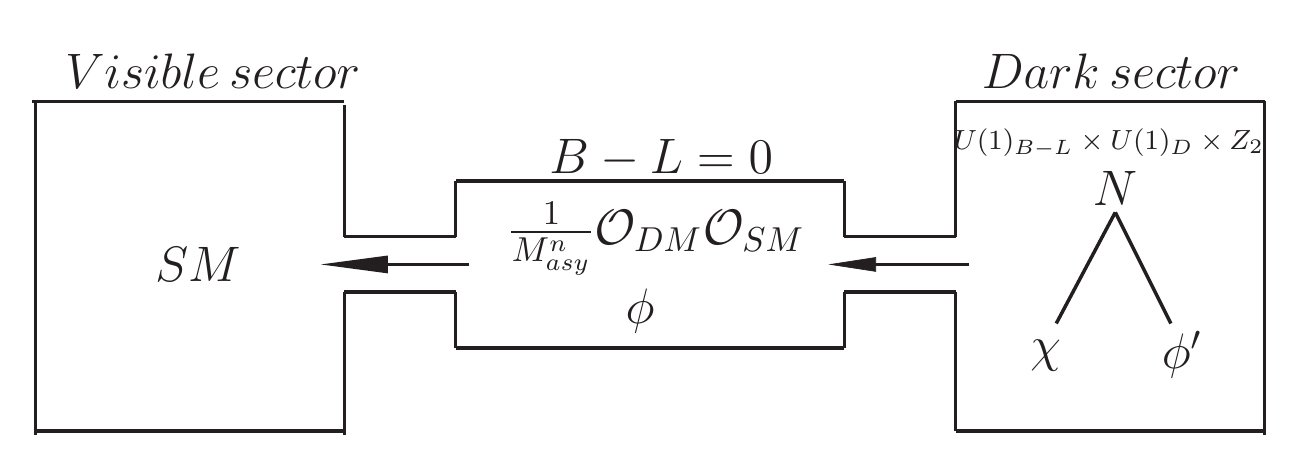}
                \caption{\footnotesize{Pictorial presentation of a dark sector being in thermal contact with the visible 
                 sector via Higgs portal coupling as well as higher dimension operators, which conserve $B-L$ symmetry and are in thermal 
                 equilibrium above sphaleron decoupling temperature.}}
                \label{asydm_cartoon}
\end{figure}

The standard model (SM), which is based on the gauge group $SU(3)_C\times SU(2)_L\times U(1)_{Y}$, is a successful theory 
of fundamental particles and their interactions. However, it does not explain either the DM abundance or baryon asymmetry 
of the Universe. Moreover, it can not explain the non-zero masses of active neutrinos. In this paper we make an attempt to 
solve these problems simultaneously in a beyond SM framework. We extend the SM by including a dark sector, as shown in Fig. \ref{asydm_cartoon}, which comprises of three generations of heavy right-handed neutrinos, a singlet scalar $\phi'$ and a singlet Dirac fermion $\chi$. These particles are charged under an additional symmetry: $U(1)_{B-L}\times U(1)_D\times Z_2$, while remain inert with respect to the SM gauge group. The $U(1)_{\rm B-L}$ is a gauge symmetry, which is broken spontaneously by the vacuum expectation value of an additional scalar $\phi_{\rm B-L}$ at a high scale, say around $10^{10}$ GeV, and give Majorana masses to right-handed neutrinos, while $U(1)_D$ is a global symmetry and is allowed to break softly by the higher dimension operators. Moreover, the $U(1)_D$ symmetry provides a distinction between the dark sector fermions $N_R$ and $\chi$, which are having same charge under $U(1)_{\rm B-L}\times Z_2$ symmetry. Due to $U(1)_D$ 
symmetry, both the singlet fermion ($\chi$) and scalar ($\phi'$) are stabilized and become viable candidates of dark matter.   

The Majorana mass of heavy right-handed neutrinos break $B-L$ symmetry by two units. Therefore, the CP-violating out-of-equilibrium 
decay of heavy right-handed neutrinos to $\chi \phi'$ in the early Universe generate a net $B-L$ asymmetry~\cite{Sakharov:1967dj,
fukugita&Yanagida_1986}. The latter is then transferred to the visible sector by a dimension eight operator~\cite{Kaplan:2009ag,Feng:2012jn,Ibe:2011hq}: 
$\mathcal{O}_8= \frac{1}{M_{asy}^{4}} \overline{\chi}^2(LH)^2$, which is in thermal equilibrium above sphaleron decoupling temperature. Note that 
the operator $\mathcal{O}_8$ breaks $U(1)_{D}$ symmetry softly, while conserves $B-L$ symmetry. As a result the $B-L$ asymmetry produced by 
the decay of right handed neutrinos will be distributed between the dark and visible sectors. When the DM $\chi$ decouples from the thermal bath, 
the asymmetry in the two sectors get segregated. Thus we get a net $B-L$ asymmetry in the visible sector proportional to the $B-L$ asymmetry in the dark 
sector. The $B-L$ asymmetry in the visible sector gets transferred to a net baryon ($B$) asymmetry via the sphaleron transitions, while the $B-L$ asymmetry 
in $\chi$ remains intact. The asymmetry in $\chi$ and $\phi'$ combinely give rise the present day relic density of DM. An additional singlet scalar 
$\phi$ is introduced which mixes with the SM Higgs $H$ and paves a path for annihilating the symmetric components of $\chi$ and $\phi'$. The abundance 
of the singlet scalar $\phi$ will not be present in the current universe due to its decay to standard model particles through Higgs mixing.

Note that $N_R$ is odd under the $Z_2$ symmetry. As a result it does not have a tree level coupling with left-handed lepton doublets as in the 
type-I seesaw model~\cite{type1_seesaw}. However, the dimension eight operator: $\mathcal{O}_\nu=\frac{1}{\Lambda^{4}}(\overline{N_R}LH)^2$ is allowed, 
where $\Lambda$ is the scale of symmetry breaking. As we discuss in section \ref{neutino mass}\,, this generates a Majorana mass of the light neutrinos at 
one loop level. Notice that the operator $\mathcal{O}_\nu$ also breaks the $U(1)_{D}$ symmetry softly.

The paper is organized as follows. In sec.~\ref{Model}\,, we introduce the model part. Sec.~\ref{neutino mass}\, is devoted to 
explain the neutrino masses. The generation of DM asymmetry is explained in sec.~\ref{Gen_Asy_DM}\,. The transfer of 
DM asymmetry to visible sector is discussed in sec.~\ref{Asy_DStoVS}\,. In sec. \ref{symmetric_ann}\,, we describe the condition 
for annihilation of symmetric component of the DM. In sec.~\ref{Pheno_sym}\,, we demonstrate the constraints on model parameters 
from invisible Higgs decay, signal strength of a SM-like Higgs, the requirement of correct relic abundance of DM and its direct 
detection. We conclude in 
sec.~\ref{conclusion}\,. 

\section{The Model} \label{Model}
The model under consideration is based on the symmetry: $SM \times U(1)_{B-L}\times U(1)_{D}\times Z_{2}$, where 
$U(1)_{B-L}$ is a local gauge symmetry and is broken spontaneously at a high scale by the vacuum expectation value of a 
singlet scalar $\phi_{\rm B-L}$ , where as $U(1)_{D}$ is a global symmetry and is allowed to break softly by higher dimensional 
operators as we discuss below. In addition to that we extend the SM particle content by introducing a dark sector which comprises of three 
generations of heavy right handed neutrinos $N_{iR}$, $i=1,2,3$, a Dirac fermion $\chi$, and a singlet scalar $\phi'$. An additional singlet 
scalar $\phi$ is also introduced, which mixes with the SM Higgs $H$. The particle content of the model, along with the quantum numbers, are 
given in the Table~\ref{tab:1}\,. Under the discrete symmetry $Z_{2}$, which remains unbroken, both $N_R$ and $\chi$ particles are odd. As a 
result the lightest $Z_2$ odd particle $\chi$ is stable and is a viable candidate of DM. In addition to that we assume $<\phi'> = 0$. This 
implies $\phi'$ is also stable due to $U(1)_D$ symmetry. As a result the relics of $\chi$ and $\phi'$ constitute the DM content of the present 
Universe.

\begin{table}
\begin{center}
\begin{tabular} {|c c|c|c|c|c|c|c|}
\hline
& Fields & $SU(3)_C$ & $SU(2)_L$ & $U(1)_Y$ & $U(1)_{B-L}$ & $U(1)_D$ & $Z_{2}$\\ 
\hline
& $N_R$ & 1 & 1 & 0 & -1 & 1 & - \\
& $\chi$ & 1 & 1 & 0 & -1 & 1/3  & - \\
& $\phi$ & 1 & 1 & 0 & 0 & 0  & +\\
& $\phi'$ & 1 & 1 & 0 & 0 & 2/3  & + \\
& $\phi_{B-L}$ & 1 & 1 & 0 & +2 & -2  & + \\
[1mm]
\hline
\end{tabular}
\end{center}
\caption{\footnotesize{Particles of the dark sector and their quantum numbers under the imposed symmetry.}}
\label{tab:1}
\end{table}

The corresponding Lagrangian can be given as:
\begin{eqnarray}
\mathcal{L} &\supset & \overline{N_{Rj}} i \gamma^{\mu} D_{\mu} {N_{R}}_{j} + \overline{\chi} i \gamma^{\mu} D_{\mu} \chi + 
\frac{1}{2} (\partial_{\mu} \phi)(\partial^{\mu} \phi)\nonumber\\ 
& + & (\partial_{\mu}{\phi'})^{\dagger}(\partial^{\mu}{\phi'}) + (D_{\mu}{\phi_{B-L}})^{\dagger}(D^{\mu}{\phi_{B-L}}) \nonumber\\
& + & M_{\chi}\overline{\chi} \chi + \lambda_{B-L}\phi_{B-L} \overline{(N_{Ri})^{c}}N_{Rj} + \lambda_{\rm DM}
\overline{\chi} \chi \phi \nonumber\\
&+& y_{i} \overline{N_{Ri}} \chi \phi' + h.c. - V(H, \phi, \phi')
\end{eqnarray}
 where
\begin{equation*}
D_{\mu} = \partial_{\mu} + i g_{B-L} Y_{B-L} (Z_{B-L})_{\mu}
\end{equation*}
and
\begin{eqnarray}\label{potential}
V(H,\phi, \phi') &=& -\mu_{H}^{2} H^{\dagger}H + \lambda_{H}(H^{\dagger}H)^{2}+\frac{1}{2} M_{\phi}^{2}\phi^{2}\nonumber\\
&+& \frac{1}{4}\lambda_{\phi}\phi^{4} + M_{\phi'}^{2} \phi'^{\dagger}{\phi'} + \lambda_{\phi'}(\phi'^{\dagger}\phi')^{2} \nonumber\\
&+& \frac{1}{2}\lambda_{H\phi}(H^{\dagger}H)\phi^{2} + \mu_\phi \phi (H^{\dagger}H) + \mu'_{\phi} \phi (\phi'^{\dagger}\phi')\nonumber\\
&+& \lambda_{H \phi'}(H^{\dagger}H)(\phi'^{\dagger}\phi') + \frac{\lambda_{\phi \phi'}}{2} \phi^{2} (\phi'^{\dagger} \phi') \,. 
\end{eqnarray}
In Eq. \ref{potential}, we assume that $U(1)_{\rm B-L}$ gauge symmetry is broken spontaneously by the vev of $\phi_{\rm B-L}$ 
at a high scale: $\langle \phi_{\rm B-L}\rangle = v_{\rm B-L} \sim 10^{10}$ GeV (say). Therefore, $\phi_{\rm B-L}$ does not play any 
role in the low energy electroweak phenomenology. However, the vev of $\phi_{\rm B-L}$ gives super heavy masses to right-handed 
neutrinos as well as neutral gauge boson $Z_{\rm B-L}$. The $B-L$ quantum numbers of $N_R$ and $\chi$ are same and is taken to be -1. 
However, they are distinguishable by their $U(1)_D$ quantum numbers. 

Since $B-L$ charges of all the SM fermions are known, it is straight forward to see that the uplifting of global $U(1)_{\rm B-L}$ 
symmetry of the SM to a gauged one brings in $B-L$ anomalies. In particular, the non-trivial one per family is given by~\cite{Geng:1988pr}: 
\begin{eqnarray*}
U(1)_{\rm B-L}^3  &:&  3 \left[ 2 \times \left(\frac{1}{3}\right)^3 - \left( \frac{1}{3} \right)^3- \left( \frac{1}{3} \right)^3\right] \nonumber\\
&+&  \left[ 2\times (-1)^3 - (-1)^3 \right]=-1  \nonumber\\
\end{eqnarray*}
where the number 3 in front is the color factor. This can be exactly cancelled by introducing one right-handed neutrino per family 
as we did in this model. Thus the model is anomaly free. Since $\chi$ is a vector-like fermion it does not introduce any additional 
anomaly though it is charged under $U(1)_{\rm B-L}$.   

As discussed above, the masses of heavy right handed neutrinos: $M_N >> M_W$, while the mass of $\chi$ is $M_\chi < M_W$.
The neutral gauge boson corresponding to $B-L$ symmetry acquires a large mass $M_{Z_{B-L}} >> M_{Z}$. In the following we 
discuss the $\phi-H$ mixing on which the annihilation of symmetric component of DM depends.

\subsection*{Case-I}
The electroweak phase transition occurs as the SM Higgs acquires a vacuum expectation value (vev) $v=\langle H \rangle$. 
This induces a non-zero vev to $\phi$ due to the trilinear term $\mu_\phi \phi (H^{\dagger}H)$ as given in Eq.~\ref{potential}. 
We assume that  $\langle \phi \rangle =u << v$. Then the quantum fluctuations around the minimum can be given as:
\begin{equation}
H=\begin{pmatrix} 0 \\ \frac{v+h}{\sqrt{2}} \end{pmatrix}\,\,\, {\rm and} \,\,\, \phi=u+\tilde{\phi}  
\end{equation}
By minimizing the scalar potential \ref{potential} we get the vacuum expectation values,
\begin{equation}
u=\frac{-\mu_{\phi} v^{2}}{M_{\phi}^{2}+\lambda_{H\phi}v^{2}}\,,
\end{equation}
and 
\begin{equation}
v=\sqrt{\frac{\mu_{H}^{2}-\frac{1}{2}\lambda_{H\phi}u^{2}-\mu_{\phi}u}{2\lambda_{H}}}\,.
\end{equation}
Notice that $\mu_\phi \propto u$. In the limit $u \rightarrow 0$, we recover the vev of SM Higgs
\begin{equation}
v=\sqrt{\frac{\mu_{H}^{2}}{2 \lambda_{H}}}\,.
\end{equation}
The small vev $u$ does not affect our discussions in the following sections and hence we set it to zero from 
here onwards.

\subsection*{Case-II}
\begin{figure}[h!]
				\centering
				\includegraphics[width = 70mm]{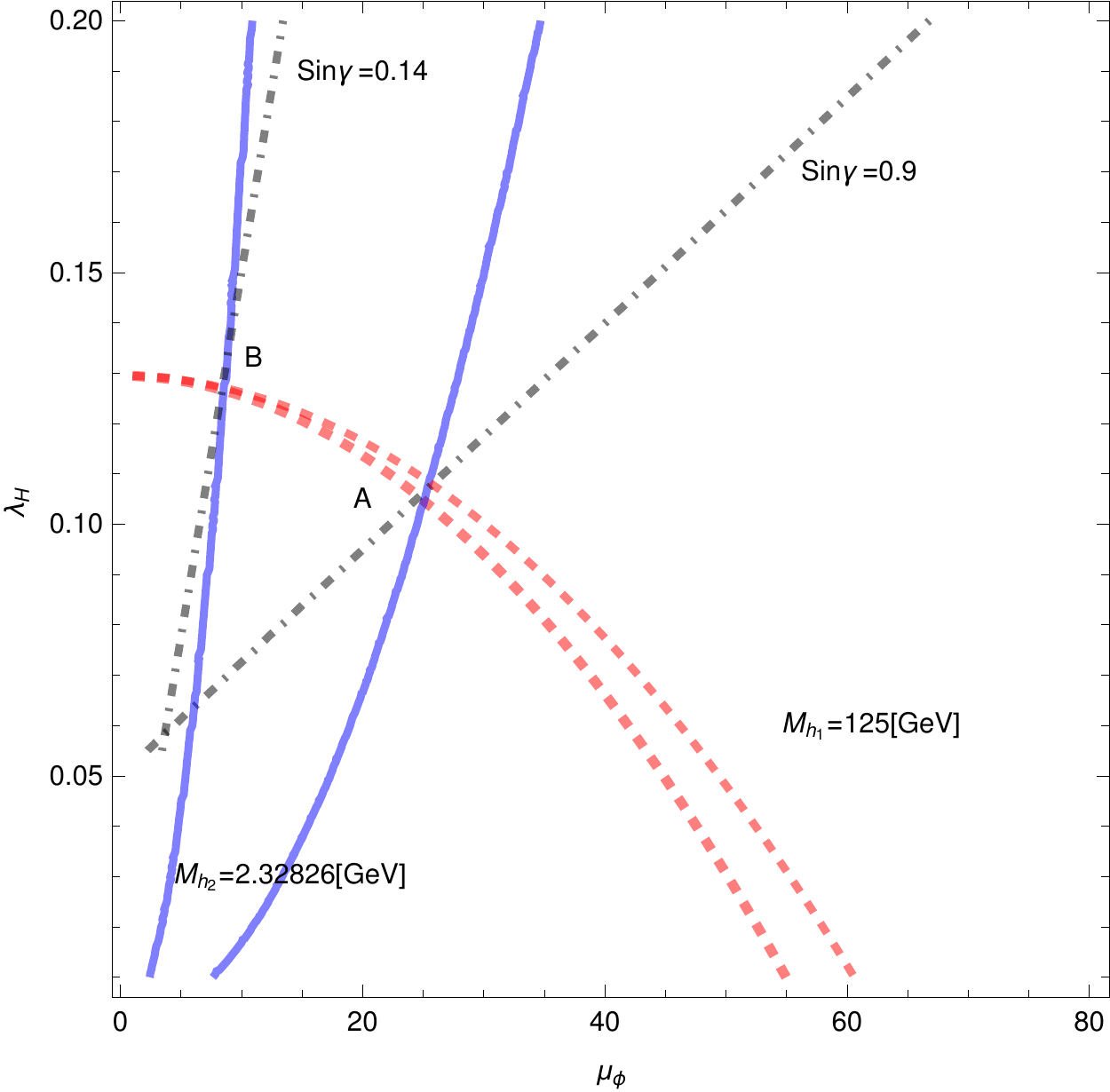}
                \caption{\footnotesize{Contours of $M_{h_1}=125.18$ GeV (dashed lines), $M_{h_2}=2.32826$ GeV (solid lines), $\sin \gamma=0.14,0.9$ 
(dot dashed lines) in the plane of $\lambda_H$ versus $\mu_\phi$. We set $M_\phi=0$.}}
                \label{higgs_mass_mixing_contour}
\end{figure}

Here we relax the $\phi$ vev to zero, i.e., $\langle \phi \rangle =0$. Then the quantum fluctuation around the minimum is given by 
\begin{equation}
H=\begin{pmatrix} 0 \\ \frac{v+h}{\sqrt{2}} \end{pmatrix}\,\,\, {\rm and} \,\,\, \phi= \tilde{\phi}\,.
\end{equation}
As a result after electroweak phase transition, the two scalars $h$ and $\tilde{\phi}$ mix with each other. The mass matrix is given by
\begin{equation}
\begin{pmatrix} 2 \lambda_H v^2  & \frac{\mu_\phi v}{\sqrt{2}}\\
\frac{\mu_\phi v}{\sqrt{2}} & M_\phi^2+ \frac{\lambda_{H\phi}}{2}v^2 \end{pmatrix}\,.
\end{equation}
Diagonalizing the above mass matrix we get the masses $M_{h_1}$ and $M_{h_2}$ corresponding to the physical Higgses $h_1$ and $h_2$:
\begin{eqnarray}
h_1 &=& h \cos \gamma + \tilde{\phi} \sin \gamma \nonumber\\
h_2 &=& - h \sin \gamma + \tilde{\phi} \cos \gamma\,.
\end{eqnarray} 
The mixing angle $\gamma$ can be quantified as
\begin{equation}
\sin \gamma \approx \frac{\sqrt{2} \mu_\phi v}{2\lambda_H v^2 - M_\phi^2 -\frac{\lambda_{H\phi} v^2}{2}}\,.
\end{equation}
We identify $h_1$ to be the SM-like Higgs with mass $M_{h_1}=125.18 {\rm GeV}$, while $h_2$ is the second Higgs whose mass is going 
to be determined from relic abundance requirement. In fact, in section \ref{Pheno_sym}\,, we obtain the light scalar mass, from the 
requirement of depletion of the symmetric component of the DM, to be  $M_{h_2} \approx 2 M_\phi' \approx 2 M_\chi\approx 
2.32826 \,{\rm GeV}$. In Fig.~\ref{higgs_mass_mixing_contour} we show the contours of $M_{h_1}=125.18$ GeV (dashed lines), 
$M_{h_2}=2.32826$\,GeV (solid lines) and $\sin \gamma=0.14, 0.9$ (dot-dashed lines) in the plane of $\lambda_H$ versus $\mu_\phi$ 
for $\lambda_{H\phi}=0.1$ (meeting at point A), $0.01$ (meeting at point B). We see that large range of mixing is allowed to explain 
simultaneously the masses of $h_1$, $h_2$. Later we will see that large mixing angles are strongly constrained by other phenomenological 
requirements. When the mixing goes to zero ({\it i.e.} $\mu_\phi \to 0$, which implies $\sin \gamma \to 0$), we recover the SM Higgs 
mass $M_{h_1} = 2 \lambda_H v^2=125.18 {\rm GeV}$ for $\lambda_H=0.13$. As the mixing angle increases ({\it i.e.} $\mu_\phi \neq 0$) we 
still satisfy the required masses of $h_1$ and $h_2$ with small $\lambda_H$. In what follows we will take $\sin \gamma$ as the measure 
of mixing.    

\begin{figure}[h!]
				\centering
				\includegraphics[width = 80mm]{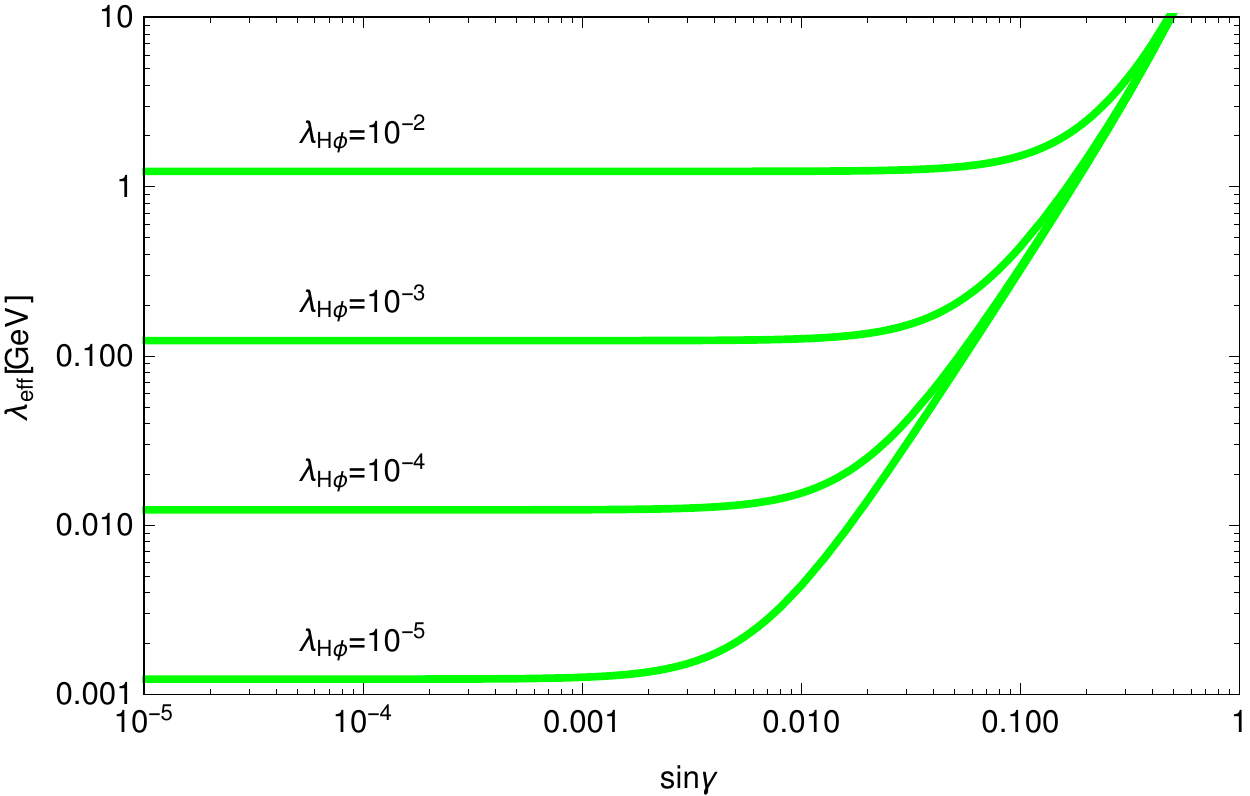}
                \caption{\footnotesize{Effective coupling of $h_1$ to $h_2 h_2$ as a function of $\sin \gamma$ for $\lambda_H=0.13$.}}
                \label{effective_h1h2h2_coupling}
\end{figure}

The effective coupling of $h_1h_2h_2$ from Eq. \ref{potential} can be given as:
\begin{eqnarray}
\lambda_{\rm eff} &=& 3 \lambda_H v \cos \gamma \sin^2 \gamma + \frac{\lambda_{H\phi}}{2} v\cos^3 \gamma +\frac{\mu_\phi}{2} \sin^3 \gamma\nonumber\\
&-& \mu_\phi \sin\gamma \cos^2\gamma -\lambda_{H\phi} v \sin^2 \gamma \cos \gamma \,. 
\end{eqnarray}
In Fig. \ref{effective_h1h2h2_coupling} we have shown the effective coupling of SM-like Higgs to $h_2 h_2$ as a function of $\sin \gamma$ 
for various values of $\lambda_{H\phi}$. We see that $\lambda_{\rm eff}$ is almost independent of $\lambda_{H\phi}$ for $\sin\gamma \sim 0.1$. 
We will come back to this issue while calculating the invisible decay width of SM-like Higgs in section \ref{Pheno_sym}\,.   

\subsection{Neutrino masses}\label{neutino mass}
\begin{figure}[h!]
				\centering
				\includegraphics[width = 60mm]{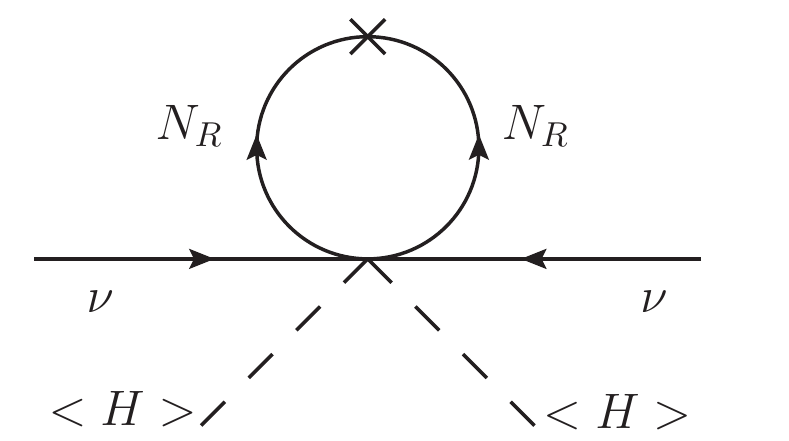}
                \caption{\footnotesize{Radiative neutrino mass at one loop level, generated by the operator $\mathcal{O}_\nu$.}}
                \label{radiative_neutrino_mass}
\end{figure}

The lepton number is violated by the Majorana mass term of the heavy right handed neutrinos. Note that the term: $\overline{N_{R}}\tilde{H}^{\dagger}L$ is 
not allowed as $N_R$ is odd under the $Z_{2}$ symmetry. However the dimension eight operator:  $\mathcal{O}_\nu= \frac{(\overline{N_{R}}\tilde{H}^{\dagger}L)^{2}}
{\Lambda^{4}}$ is allowed, where $\Lambda$ is the scale of symmetry breaking. The relevant diagram generating neutrino masses radiatively is shown in Fig.~\ref{radiative_neutrino_mass}~\footnote{See for a recent review \cite{Cai:2017jrq}.}.

By taking $\Lambda$ as the cut-off scale, the neutrino mass can be calculated from Fig. \ref{radiative_neutrino_mass} as:
\begin{equation}
M_{\nu} = \frac{1}{8 \pi^{2}}\frac{v^{2}M_{N}}{\Lambda^{2}} \left[1 + \frac{M_{N}^2}{\Lambda^{2}} Log\left(\frac{M_{N}^{2}}{M_{N}^{2}
+ \Lambda^{2}}\right) \right]\,,
\end{equation}
Where the $v$ is the vacuum expectation value of SM Higgs and $M_{N}$ is the mass scale of the heavy right-handed neutrino. Inverting 
the above formula we get the symmetry breaking scale:
\begin{equation}
\Lambda \approx 7.66 \times 10^{11} GeV \left(\frac{0.1 eV}{M_{\nu}}\right) \left(\frac{M_{N}/\Lambda}{0.1}\right).
\end{equation}
In section \ref{Gen_Asy_DM}, we take the Majorana mass of heavy right handed neutrinos to be  $M_{N} \approx 10^{10}$ GeV.

\section{Generation of asymmetry in Dark matter sector}\label{Gen_Asy_DM}
In the early Universe, the right-handed neutrinos at a temperature above their mass scales are assumed to be in thermal 
equilibrium. As the Universe expands, the temperature falls. As a result the right-handed neutrinos, below their mass scales, go 
out-of-equilibrium and decay through the process: $y_{i} \overline{N_{Ri}} \chi \phi' + {\rm h.c.}$. Without loss of 
generality we choose the mass basis of right-handed neutrinos to be diagonal. In this basis, the heavy Majorana neutrinos are 
defined by $N_i=\frac{1}{\sqrt{2}}[N_{iR} + (N_{iR})^c]$ and hence their decay violate $B-L$ by two units. In the mass basis 
of $N_1$, the decay rate of $N_1$ is given by: $\Gamma_1 = \frac{(y^\dagger y)_{11} }{16 \pi} M_1$. Comparing it with the 
Hubble expansion parameter $H=1.67 g_*^{1/2}T^2/M_{\rm Pl}$ at $T \sim M_1$, we get the out-of-equilibrium condition: 
$y \lesssim {\cal O}(10^{-3}) \sqrt{M_1/10^{10} {\rm GeV}}$. Thus depending on the mass of right-handed neutrinos the decoupling 
epoch can be different. We assume a normal hierarchy among the heavy Majorana neutrinos. As a result, the CP-violating decay of 
lightest heavy neutrino ($N_{1}$) to $\phi'$ and $\chi$, generates a net asymmetry in $\chi$ and $\phi'$. Since both $\chi$ and 
$\phi'$ are stable, the asymmetry in $\chi$ and $\phi'$ together represents the DM abundance.  

\begin{figure}[h!]
				\centering
				\includegraphics[width = 40mm]{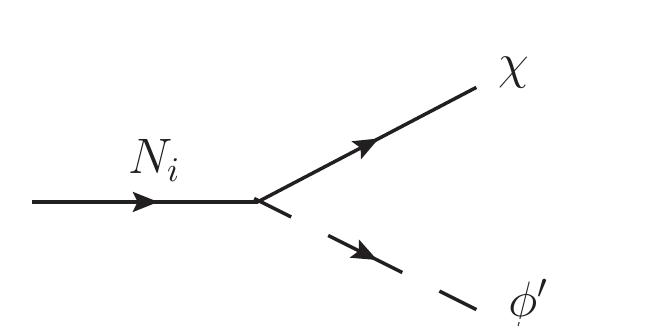}
				\includegraphics[width = 40mm]{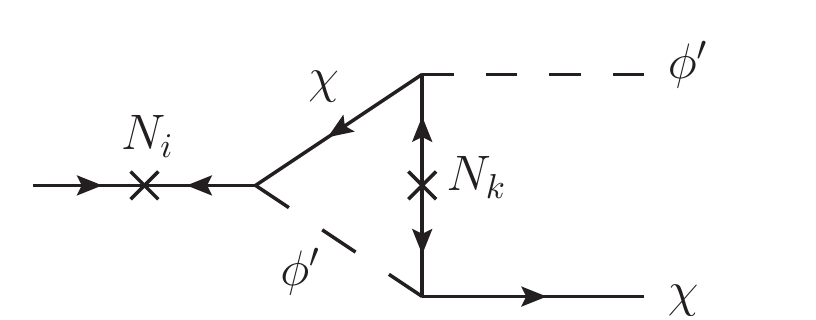}
				\includegraphics[width = 70mm]{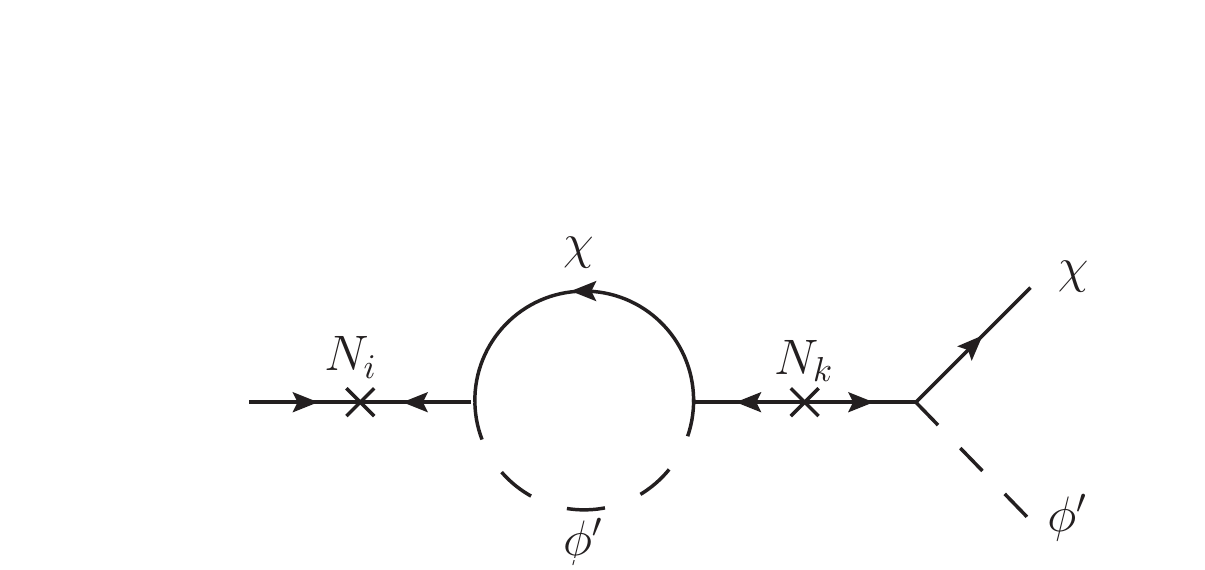}
                 \caption{\footnotesize{CP-violation arising from the interference of tree level diagram with vertex and self energy correction diagrams in the 
decay of $N_1$}}
\label{self_energy}
\end{figure}

The CP asymmetry in the decay of $N_{1}$ arises via the interference of tree level diagram with one loop self energy and vertex diagrams 
as shown in Fig.~\ref{self_energy}\,.
The asymmetry $\epsilon_\chi$ is estimated to be~\cite{buchmuller&plumacher} 
\begin{eqnarray}
\epsilon_{\chi} &=& \frac{\Gamma(N_{1}\rightarrow \chi \phi') - \Gamma (N_{1} \rightarrow \bar{\chi} \phi')}{\Gamma_{N_{1}}}\nonumber\\
               &\simeq & -\frac{1}{8\pi} \left( \frac{M_{1}}{M_{2}} \right) \frac{ {\rm Im} \left[ (y^\dagger y)^2 \right]_{12} } { (y^\dagger y)_{11} } \,.
\end{eqnarray}
where we assume $M_1 << M_2 << M_3$, and $M_i, i=1,2,3$ are the masses of heavy right-handed neutrinos. Thus below the mass scale of 
$N_1$, we get a net $B-L$ asymmetry:\cite{Buchmuller:2004nz}\cite{Giudice:2003jh}

\begin{equation}
(n_{B-L})_{total} = \epsilon_{\chi} \kappa s \times \frac{n_{N_{1}}^{eq} (T \rightarrow \infty )}{s}
\label{nb-ltotal_final}
\end{equation}
where $(n_{N_{1}}^{eq}/s)(T \rightarrow \infty)=135 \zeta(3)/(4 \pi^{4} g_{*})$ is the relativistic equilibrium abundance of $N_{1}$ . $\kappa$ is the washout factor, arises via inverse decay and scattering processes and  $s=(2\pi^2/45)g_* T^3$ is the entropy density. 
Depending on the strength of Yukawa coupling, the value of $\kappa$ can vary between 0 to 1. However, for definiteness we choose 
$\kappa = 0.01$. The details of $B-L$ asymmetry generated in the dark sector can be obtained by solving the required Boltzmann 
equations~\cite{Buchmuller:2004nz}, which is beyond the scope of this paper. The generated $B-L$ asymmetry will be 
distributed between visible and dark sectors via a higher dimensional operator as we introduce in section~\ref{Asy_DStoVS}\,.

\section{Asymmetry transfer from dark sector to visible sector} \label{Asy_DStoVS}
The asymmetry generated via the decay of lightest heavy Majorana neutrino $N_1$ can be transferred to the visible sector by 
a higher dimensional operator~\cite{Feng:2012jn}: 
\begin{equation}
\mathcal {O}_8= \frac{1}{M_{asy}^{4}} \overline{\chi}^{2}(LH)^{2}.
\label{DLAsy_dim8}
\end{equation}
Depending on the value of $M_{asy}$, the transfer operator will decouple from thermal plasma at different temperatures. We 
can find the decoupling temperature by comparing the interaction rate of the transfer operator with the Hubble expansion 
rate of the universe at the decoupling epoch $T_D$. For the operator \ref{DLAsy_dim8}, the rate of interaction between 
visible and dark sector at the decoupling epoch $T_D$ is given as 
\begin{equation}
\Gamma_{\rm D} \simeq \left( \frac{T_D^4}{M_{\rm asy}^4}\right)^2 T_D\,,
\end{equation}
where $M_{\rm Pl}$ is the Planck mass. By comparing the above interaction rate with the Hubble expansion parameter 
$H=1.67 g_*^{1/2}T_D^2/M_{\rm Pl}$ we get 
\begin{equation}\label{asy_constraint}
M_{asy}^{8} > M_{Pl} T_D^{7}\,.
\end{equation}
We assume that $T_D \gtrsim T_{\rm sph}$, where  $T_{\rm sph}$ is the sphaleron decoupling temperature. For Higgs mass $M_{h_1}=125.18$ GeV, the 
sphaleron decoupling temperature is $T_{\rm sph} \geq M_{W}$. As a result from Eq.~\ref{asy_constraint}\, we get the constraint on $M_{\rm asy}$ 
to be $M_{\rm asy} = 0.9\times10^{4}{\rm GeV}= M_{\rm asy}^*$ for $T_D = M_W$. In other words Eq. \ref{asy_constraint} indicates that, if  $M_{\rm asy}
 > M_{\rm asy}^*$, then the interaction rate of transfer operator will be in thermal equilibrium for $T > T_D$. The same condition also implies that 
the processes allowed by the operator will remain out-of-equilibrium below electroweak phase transition. Notice that the estimation of Eq. \ref{asy_constraint} 
holds only for the case where $\chi$ mass is much smaller than $T_D$. However, if one were to study heavier $\chi$, chemical decoupling can take place 
when the number density of $\chi$ becomes Boltzmann suppressed. See for instance~\cite{Bernal:2016gfn}. 

The asymmetry in the equilibrium number densities of particle $n_{i}$ and antiparticle $\overline{n_{i}}$ can be given as
\begin{equation}
n_{i}-\overline{n_{i}} = \frac{g_{i}}{2 \pi^{2}} \int_{0}^{\infty}dq q^{2} \left[ \frac{1}{e^{\frac{E_{i}(q)-\mu_{i}}{T}}\pm 1}-\frac{1}{e^{\frac{E_{i}(q)+\mu_{i}}{T}}\pm 1} \right]
\end{equation}
where the $g_{i}$ is the internal degrees of freedom of the particle species $i$. In the above equation, $E_{i}$ and $q_{i}$ are the energy 
and momentum of the corresponding particle species $i$. In the approximation of a weakly interacting plasma, where $\beta \mu_{i} \ll 1$, $\beta \equiv 1/T$, 
we get~\cite{kolbturner}
\begin{eqnarray}
n_{i}-\overline{n_{i}} & \sim & \frac{g_{i}T^{3}}{6} \times [ 2\beta\mu_{i}+\mathcal{O}((\beta \mu_{i})^{3})\,\, {\rm bosons} \nonumber\\
& \sim & \frac{g_{i}T^{3}}{6} \times [ \beta \mu_{i}+\mathcal{O}((\beta \mu_{i})^{3}) \,\,{\rm fermions}\,.
\label{particle-antiparticle-number}
\end{eqnarray}
By comparing Eq. \ref{particle-antiparticle-number} with Eq. \ref{nb-ltotal_final} we see that $\beta \mu \sim k \epsilon_\chi << 1$. This 
justifies the weak interaction of thermal plasma. We will comeback to this issue at the end of this section. 

Now we will estimate the $B$ asymmetry in the visible sector at a temperature above the sphaleron decoupling temperature. To find that we 
will use the chemical equilibration~\cite{Harvey:1990qw} between different fermions until sphaleron decoupling temperature as discussed 
below. All the left handed charged lepton $e_{iL}, \forall i$, right handed charged lepton $e_{iR}, \forall i$, left 
handed neutrino $\nu_{iL}, \forall i$, left handed up type quark $u_{iL}, \forall i$, right handed up type quark $u_{iR}, \forall i$, 
left handed down type quark $d_{iL}, \forall i$, right handed down type quark $d_{iR}, \forall i$, $W^{\pm}$, $Z$-boson, photon($\gamma$), 
Higgs boson($h$) are in thermal equilibrium until sphaleron decoupling temperature. Here the index $i = 1, 2, 3$ is written for three 
generations. All the three generations up type quark have same chemical potential, all the three generations down type quark have same 
chemical potential. Similarly all the three left handed neutrinos have same chemical potential. But the three different charge leptons 
may have different chemical potential. So we omit index $i$ from chemical potential of quarks and neutrinos. The chemical potential of 
physical Higgs boson, $Z$ boson and photon are set to zero.

Below electroweak phase transition, the Yukawa interactions can be given as:
\begin{eqnarray}
\mathcal{L}_{Yukawa} &=& g_{e_{i}}\bar{e}_{iL}he_{iR}+g_{u_{i}}\bar{u}_{iL}hu_{iR}\nonumber\\
&+&g_{d_{i}}\bar{d}_{iL}hd_{iR} +h.c\,,
\label{b1}
\end{eqnarray}
which gives the following chemical potential condition,
\begin{equation}
0=\mu_{h}=\mu_{u_{L}}-\mu_{u_{R}}=\mu_{d_{L}}-\mu_{d_{R}}=\mu_{e_{iL}}-\mu_{e_{iR}}.
\label{b2}
\end{equation}
Thus we see that for quark and charge leptons the left handed and right handed fields have same chemical potential. 
Sphaleron transitions are efficient down to the decoupling temperature $T_{\rm sph}$ and hence we get,
\begin{equation}
\mu_{u_{L}}+2\mu_{d_{L}}+\mu_{\nu}=0
\label{b6}
\end{equation}

At a temperature below electroweak phase transition the electric charge neutrality of the Universe holds. However, at this epoch the top 
quark is already decoupled from the thermal plasma and hence does not take part in the charge neutrality condition. Therefore, we get
\begin{equation}
Q=4(\mu_{u_{L}}+\mu_{u_{R}})+6\mu_{W}-3(\mu_{d_{L}}+\mu_{d_{R}})-\sum\limits_{i=1}^3(\mu_{e_{iL}}+\mu_{e_{iR}})=0\,.
\label{n1}
\end{equation}
The charge current interactions:
\begin{equation}
\mathcal{L}_{int}^{(W)}=gW_{\mu}^{+}\bar{u}_{L}\gamma^{\mu} d_{L} + gW_{\mu}^{+}e_{iL}\gamma^{\mu}\bar{\nu}_{e_{iL}}.
\label{b3}
\end{equation}
are also in thermal equilibrium below electroweak phase transition down to sphaleron decoupling temperature and hence 
satisfies the following chemical equilibrium condition:
\begin{equation}
\mu_{W}=\mu_{u_{L}}-\mu_{d_{L}},
\label{b4}
\end{equation}
\begin{equation}
\mu_{W}=\mu_{\nu}-\mu_{e_{iL}}\,, \forall i.
\label{b5}
\end{equation}
Thus Eq.~\ref{b5} ensures that three generations of charge leptons also have same chemical potential.

Thus solving the Eqs~\ref{b2}- \ref{b5}, we get the total baryon and lepton number densities in the visible sector:
\begin{equation}
n_{B} = -\frac{90}{29}\mu_{\nu}\,\,\, {\rm and} \,\,\, n_{L} = \frac{201}{29}\mu_{\nu}\,.
\label{a6}
\end{equation}
In Eq.~\ref{a6} we have dropped the common factor $ \frac{g_{i}T^{3}}{6} \times \beta$ and follow the same notation through out the draft as we 
are interested in ratio of densities, rather than their individual values. From the above Eq. \ref{a6} we get the total $B-L$ asymmetry 
in the visible sector $n_{B-L}$:
\begin{equation}
(n_{B-L})_{\rm vis} = -\frac{291}{29}\mu_{\nu}\,.
\label{a7}
\end{equation}
Moreover, from Eqs. (\ref{a6}) and (\ref{a7}) we get the total baryon asymmetry: 
\begin{equation}
n_{\rm B}=\frac{30}{97} (n_{\rm B-L})_{\rm vis}
\label{b-asymmetry}
\end{equation} 

We assume that, the dark matter $\chi$ is also in thermal equilibrium with the visible sector 
via the dimension eight operator $\mathcal{O}_8$ until the sphaleron decoupling temperature $T_{\rm sph} > M_{W}$. This gives 
chemical equilibrium condition: 
\begin{equation}
-\mu_{\chi} + \mu_{\nu} = 0
\label{a8}
\end{equation}
Thus from Eqs. \ref{a7} and \ref{a8} we get the number density of $\chi$ asymmetry, which is also the $B-L$ number density in the 
dark sector:
\begin{equation}
n_\chi = (n_{\rm B-L})_{\rm dark} = -2\mu_{\chi} = \frac{58}{291}(n_{B-L})_{\rm vis}\,.
\label{a9}
\end{equation}

The total $n_{B-L}$ of the Universe, generated by the CP-violating out-of-equilibrium decay of lightest right handed neutrino ($N_1$), is the sum of $n_{B-L}$ in the visible and dark sectors. Therefore, we get 
\begin{eqnarray}
(n_{B-L})_{\rm total} &=& (n_{B-L})_{\rm vis} + (n_{B-L})_{\rm dark}\nonumber\\
 &=& (n_{B-L})_{\rm vis} + \frac{58}{291}(n_{B-L})_{\rm vis} \nonumber\\
&=& \frac{349}{291}(n_{B-L})_{\rm vis}.
\label{a11}
\end{eqnarray}
Comparing Eq.~\ref{a11} with Eq.~\ref{nb-ltotal_final} and using Eq.~\ref{b-asymmetry} we get the required asymmetry for observed DM abundance $\epsilon_\chi = 141.23 (\eta/\kappa) (s/n_{N_{1}}^{eq} (T \rightarrow \infty ))$. Thus for $\kappa \sim 0.01$ we get $\epsilon_\chi \sim 10^{-6}$. This is in accordance with the weakly interacting plasma with $\beta \mu \sim \epsilon_\chi \approx 10^{-6}$.  Using Eq.~\ref{a11} in Eq.~\ref{b-asymmetry} and Eq.~\ref{a9}, we can get,
\begin{equation}
n_{\rm B}=\frac{90}{349} (n_{\rm B-L})_{\rm total} \,\,\,\,,\,\,\,\, n_\chi = \frac{58}{349}(n_{B-L})_{\rm total}
\label{nchi_nB_total}
\end{equation}

The Asymmetry generated in $\phi'$ can be written as,
\begin{equation}
n_{\phi'} = (n_{B-L})_{\rm total}
\label{nphi'_total}
\end{equation}

\begin{figure}[h!]
				\centering
				\includegraphics[width = 80mm]{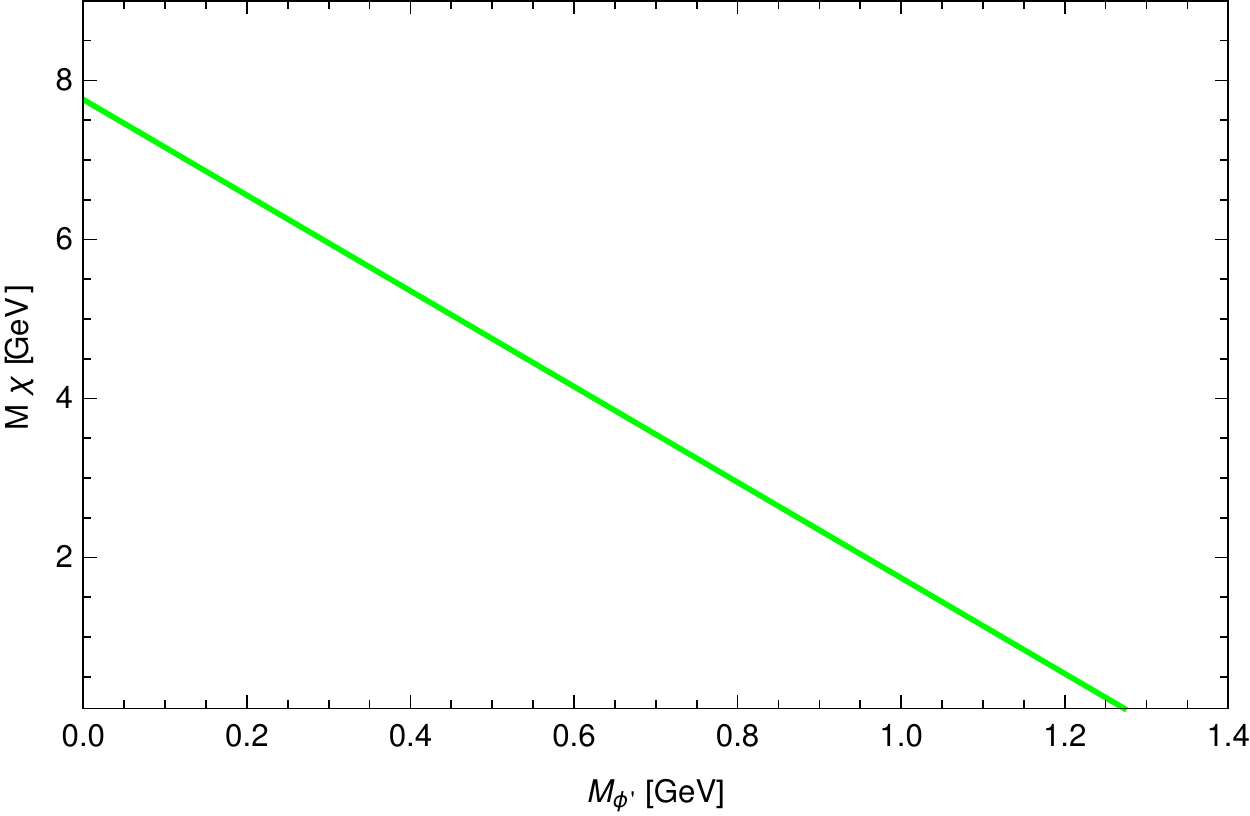}
                \caption{\footnotesize{Allowed mass range for both DM candidates.}}
                \label{mphip_mchi}
\end{figure}

The ratio of DM to baryon abundance, given by WMAP and the PLANCK data, to be $\Omega_{\rm DM}/ \Omega_{\rm B} \approx 5$. This implies from   
Eqs. (\ref{nchi_nB_total}) and (\ref{nphi'_total}):
\begin{equation}
M_{\chi} = \frac{450 M_{p} - 349 M_{\phi'}}{58}
\end{equation}
where $M_{p}$ is the proton mass. We have shown the allowed range of masses of $\chi$ and $\phi'$ in Fig. \ref{mphip_mchi}. 
In what follows, we take $M_{\chi}=M_{\phi'} \approx 1$ GeV.

\section{Annihilation of the symmetric component of the dark matter}\label{symmetric_ann}
\begin{figure}[h!]
				\centering
				\includegraphics[width = 80mm]{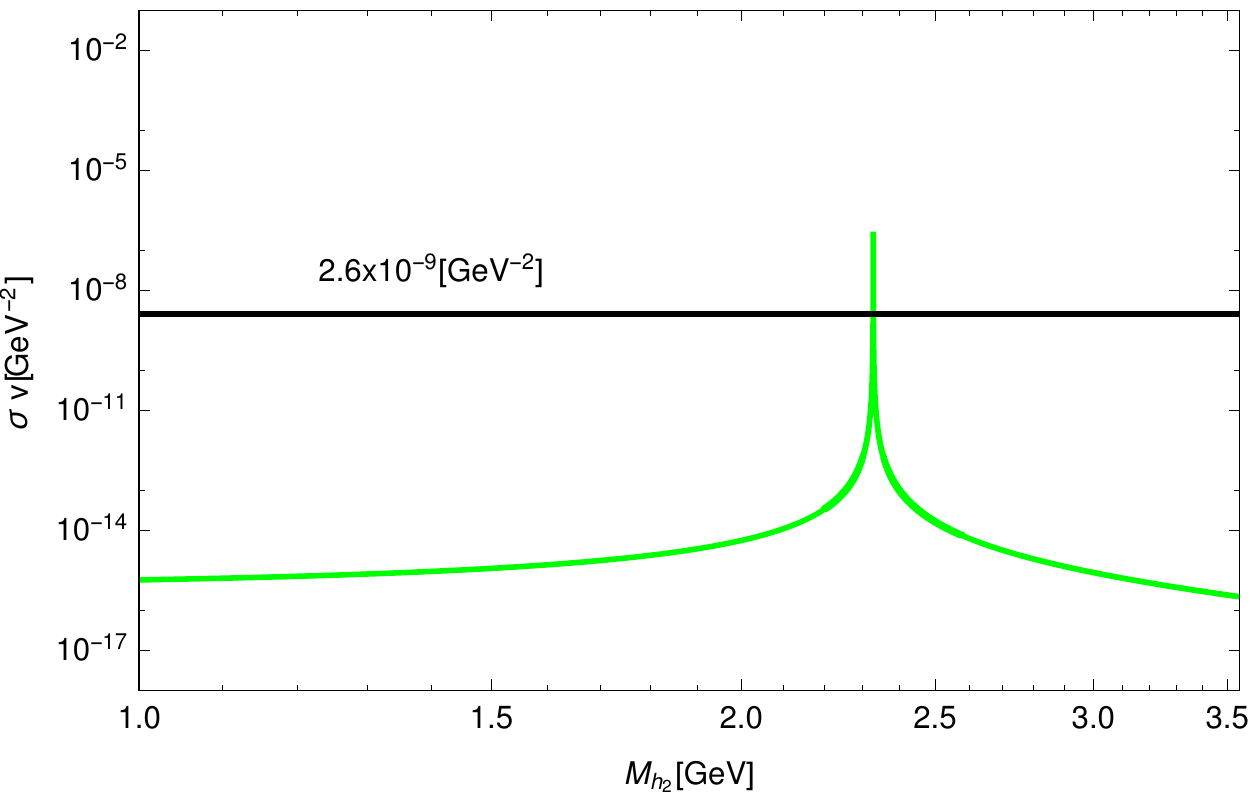}
					 \caption{\footnotesize{The annihilation cross-section of $\overline{\chi}\chi \rightarrow \overline{f}f$ as 
a function of $M_{h_{2}}$ for a typical value of $\lambda_{\rm DM}=1\times 10^{-2}$, $\lambda_{\rm H\phi'}=1\times 10^{-3}$, $\mu_{\phi'}=1 \times 10^{-3}$ and  $\sin \gamma=0.1$.}}
\label{depletion_chi}
\end{figure}

\begin{figure}[h!]
				\centering
				\includegraphics[width = 80mm]{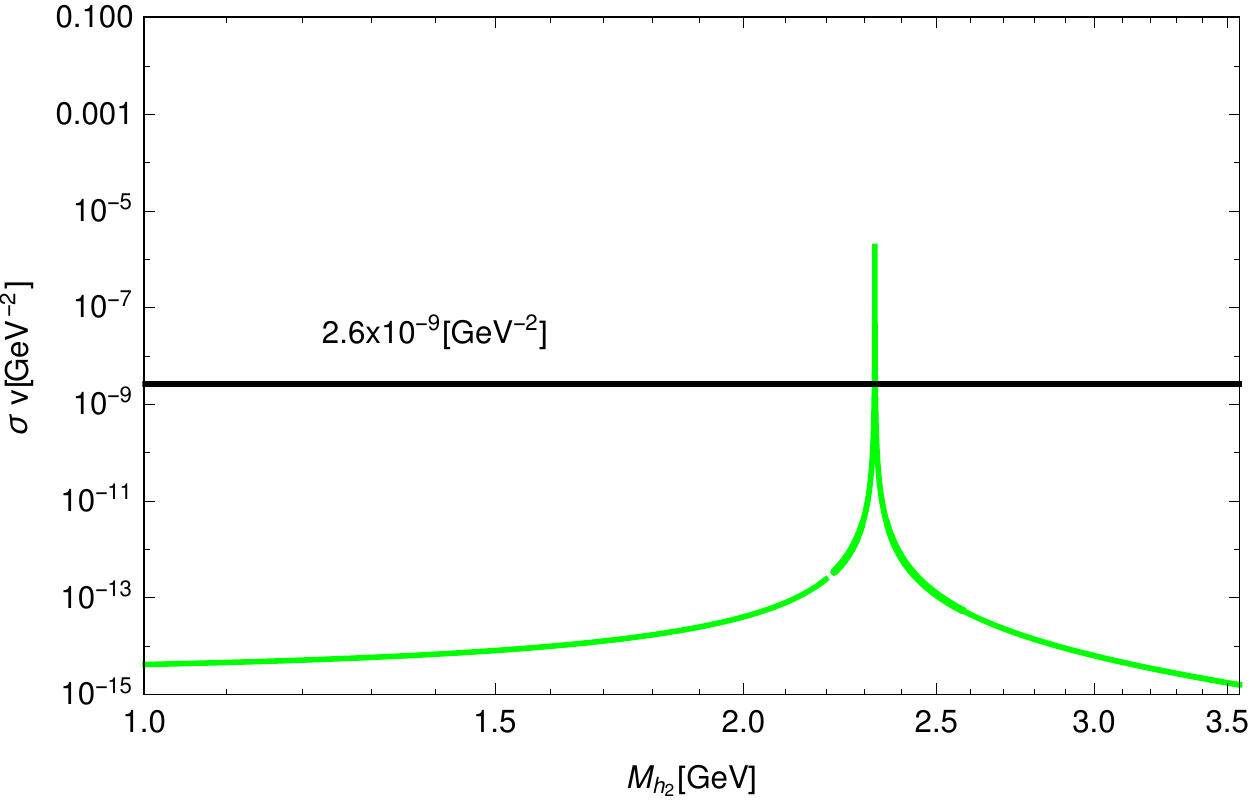}
					 \caption{\footnotesize{The annihilation cross-section of $\overline{\phi^{\dagger}}\phi' \rightarrow \overline{f}f$ as 
a function of $M_{h_{2}}$ for a typical value of $\lambda_{\rm DM}=1\times 10^{-2}$, $\lambda_{\rm H\phi'}=1\times 10^{-3}$, $\mu_{\phi'}=1 \times 10^{-3}$ and  $\sin \gamma=0.1$.}}
\label{depletion_phip}
\end{figure}

The symmetric component of $\chi$ and $\phi'$ can be efficiently depleted through the $\phi$ mediated 
interactions. In particular, $\phi-H$ mixing provides a portal for annihilation of $\chi$ and $\phi'$ to the 
SM particles. We show that when the extra scalar mass ($M_{h_2})$ is twice of the DM mass we get Breit-Wigner 
enhancement in the cross-section which actually annihilates the symmetric component of the DM candidates, as 
shown in Figs.~\ref{depletion_chi}\,,~\ref{depletion_phip}\,.

The annihilation cross-section for the process: $\overline{\chi}\chi \to \overline{f} f$ is given by:
\begin{eqnarray}
\sigma_{\chi} v &=& \frac{\sqrt{s-4M_{f}^{2}}}{16 \pi s \sqrt{s}}  \nonumber\\
& \times & \frac{\lambda_{DM}^{2} \lambda_f^{2} \cos^{2}\gamma \sin^{2}\gamma}{\left[(s-M_{h_{1}}^{2})^{2}+\Gamma_{h_{1}}^{2}M_{h_{1}}^{2} \right]\left[(s-M_{h_{2}}^{2})^{2}+\Gamma_{h_{2}}^{2}M_{h_{2}}^{2} \right]} \nonumber\\
&\times & \lbrace \left[2 s-(M_{h_{1}}^{2}+M_{h_{2}}^{2}) \right]^{2} + \left[\Gamma_{h_{1}}M_{h_{1}}+\Gamma_{h_{2}}M_{h_{2}} \right]^{2} \rbrace \nonumber \\
& \times & \lbrace (s-2M_{\chi}^{2})(s-2M_{f}^{2})-2M_{f}^{2}(s-2M_{\chi}^{2}) \nonumber\\
& - & 2M_{\chi}^{2}(s-2M_{f}^{2})+4 M_{\chi}^{2} M_{f}^{2}\rbrace 
\label{scattering_cross_chi}
\end{eqnarray}
where $M_{f}$ represents the mass of SM fermions and $\lambda_f=M_f/v$. The decay width of $h_1$ is given by:
\begin{equation}
\Gamma_{h_1}=\cos^2 \gamma \Gamma_{h_1}^{SM} + \sin^2 \gamma \Gamma_{h_1}^{\bar{\chi} \chi} + \Gamma_{h_1}^{h_2h_2} + \Gamma_{h_1}^{\phi'^{\dagger} \phi'}\,,
\label{gamma_h1}
\end{equation}
where $\Gamma^{SM}_{h_1}=4.2$ MeV,
\begin{equation}
\Gamma_{h_1}^{\bar{\chi} \chi} = M_{h_{1}}\frac{\lambda_{DM}^{2}}{8 \pi} \left[ 1-\frac{4M_{\chi}^{2}}{M_{h_{1}}^{2}}\right]^{\frac{3}{2}}
\label{h1tochichi}
\end{equation}
,
\begin{equation}
\Gamma_{h1}^{h_{2}h_{2}} = \frac{\lambda_{\rm eff}^{2} }{32 \pi M_{h_{1}}} \left[1-\frac{4 M_{h_{2}}^{2}}{M_{h_{1}}^{2}} \right]^{\frac{1}{2}}\,
\label{h1toh2h2}
\end{equation}
and
\begin{equation}
\Gamma_{h_{1}}^{\phi'^{\dagger} \phi'} = \frac{(\mu_{\phi}^{'} \sin\gamma + \lambda_{H\phi^{'}}v \cos\gamma)^{2}}{32 \pi M_{h_{1}}} 
\left[1-\frac{4 M_{\phi'}^{2}}{M_{h_{1}}^{2}} \right]^\frac{1}{2}.
\label{h1phi'phi'}
\end{equation}
The decay width of $h_2$ is given by:
\begin{eqnarray}
\Gamma_{h_2} &=& \sum_f \frac{C_f M_{h_2} \sin^2 \gamma}{8\pi} \left( \frac{M_f}{v}\right)^2 \left[1-\frac{4 M_f^2}{M_{h_2}^2}  \right]^{3/2}\nonumber\\
&+& \frac{M_{h_2} \lambda_{\rm DM}^2\cos^2 \gamma}{8 \pi} \left[1-\frac{4 M_\chi^2}{M_{h_2}^2}  \right]^{3/2}\,,
\label{gamma_h2}
\end{eqnarray}
where $C_f$ accounts the color factor of SM fermions. 

The annihilation cross-section for the process: $\phi'^{\dagger}\phi' \to \overline{f} f$ is given by:
\begin{eqnarray}
\sigma_{\phi'} v &=& 
\frac{(s-4 M_{f}^{2})^{\frac{3}{2}}}{8\pi s\sqrt{s}}\nonumber\\
&\times & [ \frac{\lambda_{1}^{'2}\lambda_{2}^{'2}}{(s-M_{h_{2}})^{2}+\Gamma_{h_{2}}^{2}M_{h_{2}}^{2}}
 + \frac{\lambda_{1}^{''2}\lambda_{2}^{''2}}{(s-M_{h_{1}})^{2}+\Gamma_{h_{1}}^{2}M_{h_{1}}^{2}} \nonumber\\
& + & 2 \lambda_{1}^{'} \lambda_{2}^{'} \lambda_{1}^{''} \lambda_{2}^{''} \nonumber\\
&\times & (\frac{(s-M_{h_{2}}^{2})(s-M_{h_{1}}^{2})
 + \Gamma_{h_{1}}M_{h_{1}}\Gamma_{h_{2}}M_{h_{2}}}{[(s-M_{h_{2}}^{2})^{2}+\Gamma_{h_{2}}^{2}M_{h_{2}}^{2}][(s-M_{h_{1}}^{2})^{2}+\Gamma_{h_{1}}^{2}M_{h_{1}}^{2}})]
 \label{scattering_cross_phi'}
\end{eqnarray}
where $\lambda_{1}^{'}= \mu_{\phi}^{'} \cos \gamma- \lambda_{H\phi^{'}}v \sin \gamma $, $\lambda_{2}^{'}=-(M_{f}/v)\sin\gamma$, $ \lambda_{1}^{''}= \mu_{\phi}^{'} \sin\gamma + \lambda_{H\phi^{'}}v \cos\gamma$ and $\lambda_{2}^{''}=(M_{f}/v)\cos\gamma$. The $\Gamma_{h_{1}}$ and $\Gamma_{h_{2}}$ are given in Eqs.~\ref{gamma_h1}\,,~\ref{gamma_h2}\,.

In our case $\overline{\chi}\chi$ is annihilating dominantly to a pair of muons. In Eqs.~\ref{scattering_cross_chi}\, and \,\ref{scattering_cross_phi'}\,, 
the unknown parameters which dominantly contribute to the annihilation cross-section are 
the mass of $h_2$, {\it i.e.} $M_{h_2}$ and the singlet-doublet Higgs 
mixing, {\it i.e.} $\sin \gamma$, the coupling of $h_2$ with $\chi$, {\it i.e.} $\lambda_{\rm DM}$ and the coupling 
of $h_2$ with $\phi'$, {\it i.e.} $\lambda_{\rm H\phi'}$. However, these parameters are strongly constrained by invisible 
Higgs decay~\cite{Khachatryan:2016whc}, relic abundance of DM measured by PLANCK~\cite{Ade:2015xua} and WMAP~\cite{Hinshaw:2012aka}, 
and spin-independent direct detection cross-sections at XENON100~\cite{Aprile:2012nq}, LUX~\cite{Akerib:2016vxi}, XENON1T~\cite{Aprile:2015uzo} and CRESST-II~\cite{Angloher:2015ewa}
 and the Higgs signal strength measured at LHC~\cite{cms_report_2018, Khachatryan:2016vau}. For a typical value of the parameters 
$\lambda_{\rm DM}=1\times 10^{-2}$, $\lambda_{\rm H\phi'}=1\times 10^{-3}$, $\mu_{\phi'}=1 \times 10^{-3}$ and  $\sin \gamma=0.1$, we have 
plotted $\sigma v$ as a function of $M_{h_2}$ in Figs.~\ref{depletion_chi}\,,~\ref{depletion_phip}\,, respectively. 

As shown in Figs.~\ref{depletion_chi}\,,~\ref{depletion_phip}\,, the value of the $\chi$ and $\phi'$ annihilation cross-section $\sigma v < \langle \sigma |v| \rangle_F=2.6\times 10^{-9}/{\rm GeV^{2}}$ in most of the parameter space except at the resonance, where $\sigma v > \langle \sigma |v|\rangle_F$. A crucial observation here 
is that mass of $h_2$ has to be twice the DM mass in order to get a large cross-section via resonance. Note that a large cross-section is required to deplete the symmetric component of the DM.


\section{Phenomenological constraints}\label{Pheno_sym}

\subsection{Higgs signal strength}
\begin{figure}[h!]
				\centering
				\includegraphics[width = 70mm]{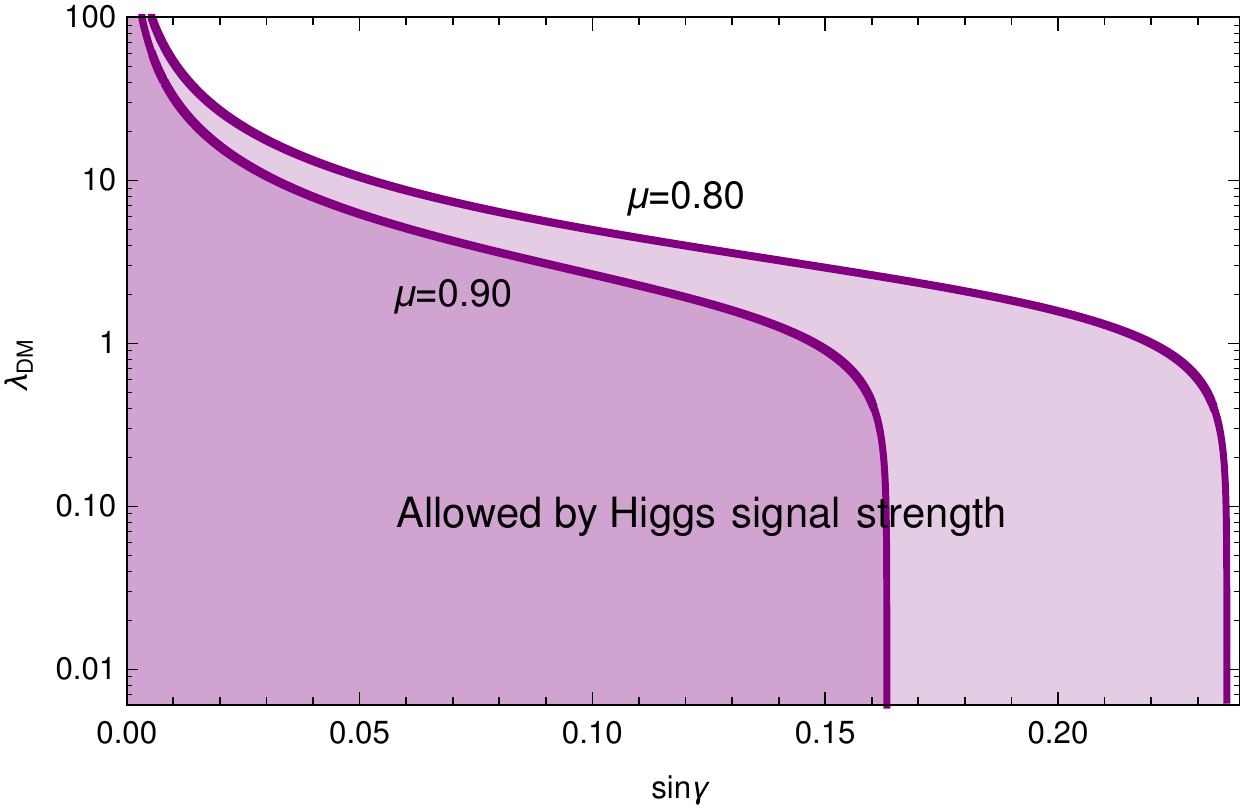}
				 \caption{\footnotesize{Contours of signal strength $\mu$ of the SM-like Higgs $h_1$ in the plane of $\lambda_{\rm DM}$ 
versus $\sin \gamma$.  }}
                \label{signal_strength_chi}
\end{figure}

\begin{figure}[h!]
				\centering
				\includegraphics[width = 70mm]{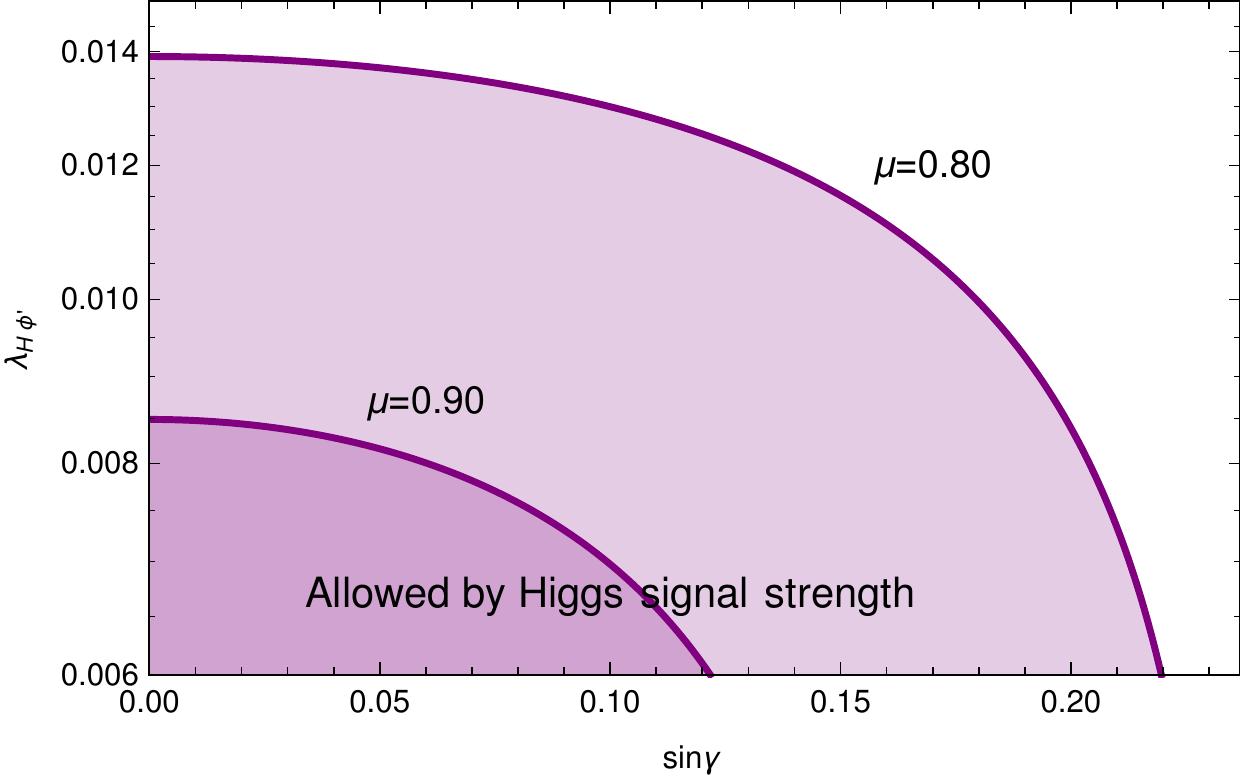}
				 \caption{\footnotesize{Contours of signal strength $\mu$ of the SM-like Higgs $h_1$ in the plane of $\lambda_{\rm H\phi'}$ versus $\sin \gamma$.}}
                \label{signal_strength_phip}
\end{figure}

The signal strength of SM-like Higgs in a particular channel $h_1 \to xx$ can be measured at LHC and can be 
defined as 
\begin{eqnarray}
\mu_{h_1 \to xx} &=& \frac{\sigma_{h_1}}{\sigma_{h_1}^{\rm sM}} \frac{{\rm Br}_{h_1 \to xx} }{{\rm Br}^{\rm SM}_{h_1 \to xx}} \nonumber\\
&=& \frac{\cos^4 \gamma \Gamma_{h_1}^{SM} }{ \Gamma_{h_1}}\,,
\end{eqnarray}
where $\Gamma_{h_1}$ is given by Eq.~\ref{gamma_h1} . 
In absence of any new physics $\mu=1$. However, in our case the mixing between the two Higgses can reduce the signal strength 
of SM-like Higgs. Therefore, the mixing can not be arbitrarily large and can be strongly constrained from the observation. The combined signal strength is measured to be $\mu=1.17 \pm 0.1$~\cite{cms_report_2018,Khachatryan:2016vau}. In Fig. \ref{signal_strength_chi}  and  \ref{signal_strength_phip} we have shown the contours of different values of $\mu$ in the planes of $\lambda_{\rm DM}$ and $\lambda_{\rm H\phi'}$ with $\sin \gamma$, respectively. From the Figs. \ref{signal_strength_chi} and \ref{signal_strength_phip} we see that as the mixing increases the signal strength reduces accordingly. For an optimistic low value $\mu=0.80$, the allowed mixing angle can be as large as 0.28. 
Thus in the rest of the draft we restrict the range of $\sin \gamma$ up to 0.3.

\subsection{Constraints from invisible Higgs decay}
The singlet-doublet Higgs mixing in this model allows the SM like Higgs $h_1$ to decay via invisible channels: 
$h_1 \to h_2 h_2$, $h_1 \to \bar{\chi}\chi$ and $h_1 \to \bar{\phi'}^{\dagger}\phi'$. The branching ratio for the invisible Higgs decay can be 
defined as: 
\begin{equation}
Br_{\rm inv} = \frac{ \sin^2 \gamma \Gamma_{h_1}^{\bar{\chi} \chi} + [{\rm Br}(h_2 \to \bar{\chi} \chi + h_2 \to \phi'^{\dagger} \phi')] \Gamma_{h_1}^{h_2h_2} + \Gamma_{h_1}^{\phi'^{\dagger}\phi'}} {\cos^2 \gamma \Gamma_{h_1}^{SM} + \sin^2 \gamma \Gamma_{h_1}^{\bar{\chi} \chi} + \Gamma_{h_1}^{h_2h_2}+ \Gamma_{h_1}^{\phi'^{\dagger}\phi'}}
\label{invisible_higgs_decay}
\end{equation}
where $\Gamma_{h_1}^{\bar{\chi} \chi}$, $\Gamma_{h_1}^{h_2h_2}$ and $\Gamma_{h_1}^{\phi'^{\dagger}\phi'}$ are given by Eqs. \ref{h1tochichi}\,,\, \ref{h1toh2h2}\, and \,\ref{h1phi'phi'}\, respectively. Note that LHC give an upper bound to the invisible Higgs decay to be $Br_{inv} \leq 24\% $~\cite{Khachatryan:2016whc}. 
For a given $h_2$ mass, the allowed invisible Higgs decay width will constraint $\lambda_{\rm DM}$, $\lambda_{\rm H\phi'}$ and $\sin \gamma$ as we discuss below.

\subsection{Constraints from direct detection of dark matter}
The singlet-doublet scalar mixing also allows the DM $\chi$ and $\phi'$ to scatter off the nucleus at terrestrial laboratories. The spin 
independent DM-nucleon scattering cross-section can be written as \cite{Goodman:1984dc} \cite{Ellis:2008hf} \cite{Akrami:2010dn} \cite{NSNS} ,
 \begin{equation}
 \sigma^{SI} = \frac{\mu_{r}^{2}}{\pi A^{2}}[Z f_{p}+(A-Z)f_{n}]^{2}
 \label{DD1}
 \end{equation}
Where the Z and A are the atomic and mass numbers of the target nucleus. In Eq. \ref{DD1}\,, the reduced mass $\mu_{r} = M_{\chi}m_{n}/(M_{\chi}+m_{n})$, 
where $m_{n}$ is the mass of the nucleon (proton or neutron) and $f_{p}$ and $f_{n}$ are the effective interaction strengths of DM with proton and 
neutron of the target nucleus and are given by:
	\begin{equation}
	f_{p,n}=\sum\limits_{q=u,d,s} f_{T_{q}}^{p,n} \alpha_{q}\frac{m_{p,n}}{m_{q}} + \frac{2}{27} f_{TG}^{p,n}\sum\limits_{q=c,t,b}\alpha_{q} 
\frac{m_{p,n}}{m_{q}}\,,
\label{DD2}
	\end{equation}
where in case of $\chi$ DM can be written as
\begin{equation}
 \alpha_{q} = \lambda_{DM} \left( \frac{m_{q}}{v}\right) \left[\frac{1}{M_{h_{2}}^{2}}-\frac{1}{M_{h_{1}}^{2}}\right] \sin\gamma \cos\gamma\,.
  \label{DD4_chi}
 \end{equation}
 where in case of $\phi'$ DM can be written as
\begin{equation}
 \alpha_{q} = \left(\frac{\mu_{\phi'}\sin\gamma + \lambda_{H \phi'}v \cos\gamma}{M_{h_{2}}^{2}}-\frac{\mu_{\phi'}\cos\gamma - \lambda_{H \phi'}v \sin\gamma}{M_{h_{1}}^{2}}\right)
  \label{DD4_phip}
 \end{equation}
In the above Eq. \ref{DD2}, the $f_{T_{q}}^{p,n}$ are given by $f_{Tu}^{(p)}=0.020\pm 0.004, f_{Td}^{(p)}=0.026\pm0.005, f_{Ts}^{(p)}=0.118\pm0.062, 
f_{Tu}^{(n)}=0.014\pm0.003, f_{Td}^{(n)}=0.036\pm0.008, f_{Ts}^{(n)}=0.118\pm0.062$ \cite{Ellis:2000ds}. The coupling of DM with the gluons in target 
nuclei is parameterized by
	\begin{equation}
	f_{TG}^{p,n}=1-\sum\limits_{q=u,d,s} f_{T_{q}}^{p,n}\,.
	 \label{DD3}
	\end{equation}
\begin{figure}[h!]
				\centering
				\includegraphics[width = 80mm]{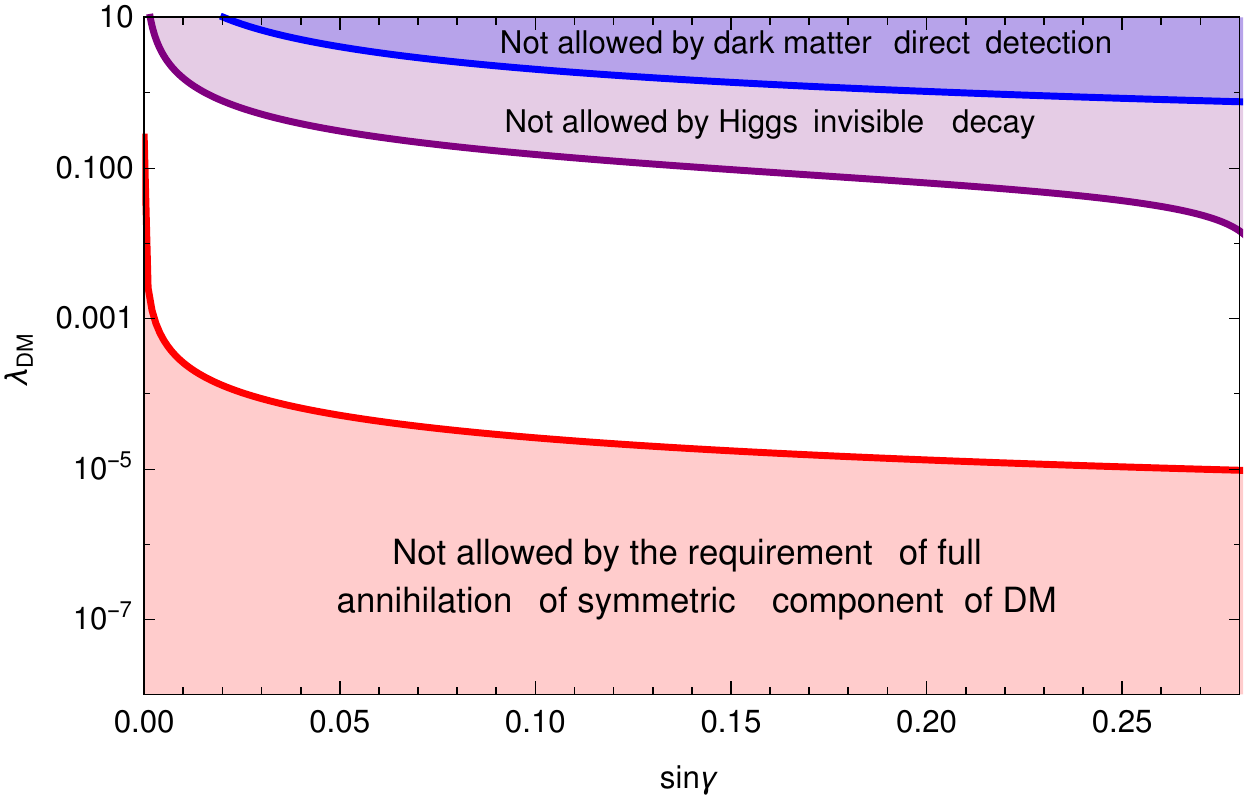}
				 \caption{\footnotesize{Allowed regions in the plane of $\lambda_{DM}$ versus $\sin \gamma$. The region above 
the top purple line is disallowed by invisible Higgs decay, {\it i.e.} $Br_{\rm inv} \geq 24 \%$. The region below the bottom Red line is disallowed because $\sigma v < 2.6 \times 10^{-9}/{\rm GeV}^2$ and give large relic abundance. The regions above the Blue line is disallowed by the spin independent direct detection cross-sections at CRESST-II 2016 for DM mass 1 GeV. we fix $M_{h_2} \approx 2 M_\chi$. }}
                \label{lambdaDM_vs_singm}
\end{figure} 

\begin{figure}[h!]
				\centering
				\includegraphics[width = 80mm]{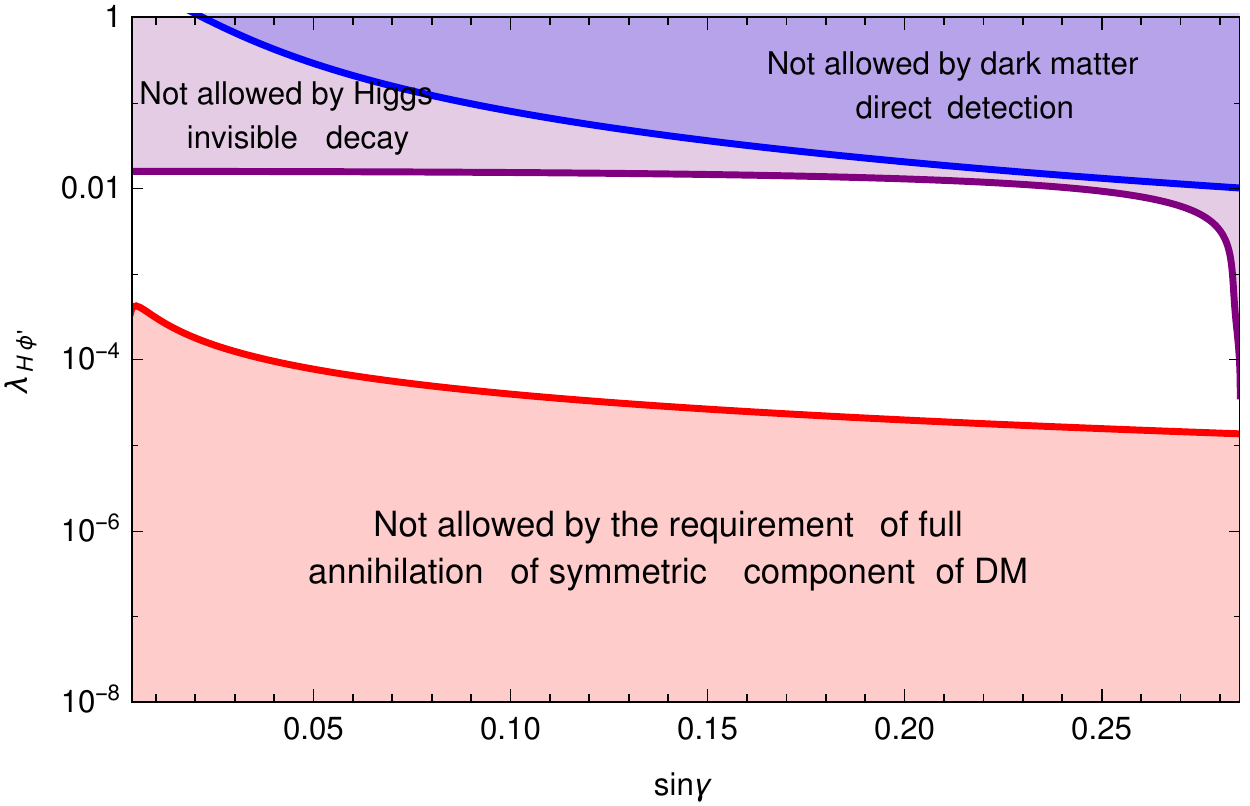}
				 \caption{\footnotesize{Allowed regions in the plane of $\lambda_{H\phi'}$ versus $\sin \gamma$. The region above 
the top purple line is disallowed by invisible Higgs decay, {\it i.e.} $Br_{\rm inv} \geq 24 \%$. The region below the bottom Red line is disallowed because $\sigma v < 2.6 \times 10^{-9}/{\rm GeV}^2$ and give large relic abundance. The regions above the Blue line is disallowed by the spin independent direct detection cross-sections at CRESST-II 2016 for DM mass 1 GeV. we fix $M_{h_2} \approx 2 M_\phi'$. }}
                \label{lambdaHphip_vs_singm}
\end{figure}

We summarize all the constraints from invisible Higgs decay, relic abundance of DM and null detection of DM at CRESST-II~\cite{Angloher:2015ewa} as 
the allowed regions in the plane of $\lambda_{\rm DM}$ versus $\sin \gamma$ in Fig.~\ref{lambdaDM_vs_singm} and $\lambda_{\rm H\phi'}$ versus 
$\sin \gamma$ in Fig.~\ref{lambdaHphip_vs_singm}. We see from Fig.\ref{lambdaDM_vs_singm} that the region above the top purple line is not 
allowed by the invisible Higgs decay, where $Br_{\rm inv} \geq 24 \%$~\cite{Khachatryan:2016whc}. The region below the bottom red line gives 
large relic abundance of DM since in this region $\sigma v (\bar{\chi} \chi \,,\, \phi'^{\dagger} \phi' \to \bar{f} f) < 2.6\times 10^{-9}/{\rm GeV}^2$. Since most of the annihilation occurs at the resonance, we fix $M_{h_2} \approx 2 M_\chi \approx 2 M_\phi'$. The Blue line 
indicates the spin independent direct detection cross-section from the CRESST-II~\cite{Angloher:2015ewa} detector results $\sigma^{\rm SI}=10^{-38} cm^{2}$  corresponding to a DM mass: $M_\chi=M_{\phi'}=1$ GeV. Therefore, the region above to that line is not allowed. Thus we left with a white allowed patch in 
the plane of $\lambda_{\rm DM}$ versus $\sin \gamma$ and $\lambda_{\rm H\phi'}$ versus $\sin \gamma$ .

\section{Conclusion}\label{conclusion}
In this paper we extended the standard model by including a dark sector which consists of three generations of heavy right-handed 
neutrinos $N_{iR}, i=1,2,3$, a singlet Dirac fermion $\chi$ and a singlet scalar $\phi'$, where the latter two particles 
represent the DM. These particles are charged under an extended symmetry $U(1)_{\rm B-L}\times U(1)_D\times Z_2$, while 
remained inert with respect to the SM symmetry. An additional singlet scalar $\phi_{\rm B-L}$ was introduced to break the 
$U(1)_{\rm B-L}$ gauge symmetry at a high scale, say $10^{10}$ GeV. The breaking of $B-L$ symmetry at a high scale not only 
gave large Majorana masses to heavy right-handed neutrinos but also made $Z_{\rm B-L}$ super heavy. The global $U(1)_D$ symmetry, 
which was softly broken by dimension eight operators, provides a distinction between $N_R$ and $\chi$ since they carry same 
charges under the $U(1)_{\rm B-L} \times Z_2$. 

In the early Universe, the CP-violating out-of-equilibrium decay of lightest heavy right handed neutrino to $\chi$ 
and $\phi'$ generate a net DM asymmetry. The latter is then transferred to visible sector via a dimension eight operator 
$(\bar{\chi}LH)^2/M_{\rm asy}^4$ which conserves $B-L$ symmetry and is in thermal equilibrium down to sphaleron decoupling temperature 
$T_{\rm sph}$. $B+L$ violating sphaleron transitions are in thermal equilibrium down to a temperature $T_{\rm sph}$ and hence can convert 
the $B-L$ asymmetry in the visible sector to a net $B$ asymmetry while the $B-L$ asymmetry in the dark sector remain untouched. As a 
result we get a net asymmetric DM abundance (given in terms of $B-L$ asymmetry) comparable to baryon asymmetry for $M_\chi\sim M_\phi'\sim M_p$, 
where $M_p$ represents proton mass. An additional light singlet scalar $\phi$ was introduced, which helped in annihilating the symmetric 
component of the DM through its mixing with the SM Higgs. We found that the efficient annihilation of the symmetric component of DM 
requires the singlet scalar mass to be around twice the DM mass irrespective of all other parameters in the model. Since 
the observed DM abundance gives the DM mass $M_\chi=M_{\phi'} \approx 1$ GeV, we get the singlet scalar mass $\approx 2$ GeV, which can 
be searched at the collider and via indirect gamma ray search. 

The neutrinos are massless at the tree level since the right handed neutrinos are odd under the $Z_2$ symmetry and are decoupled from the visible sector. 
However, at one loop level the neutrinos acquired masses via a dimension eight operator $(\overline{N_R}LH)^2/\Lambda^4$. We showed that sub-eV 
masses of neutrinos require the $B-L$ breaking scale to be around $\Lambda \approx 10^{11}$ GeV.

\section*{Acknowledgements}
SS thanks Arunansu Sil, Manimala Mitra, Kirtiman Ghosh, Pankaj Agrawal, Aruna Kumar Nayak, Debottam Das and Bhupal Dev for useful discussions.


\begin{thebibliography}{}

\bibitem{dm_review}
G.~Jungman, M.~Kamionkowski and K.~Griest,
  Phys.\ Rept.\  {\bf 267}, 195 (1996)
  [hep-ph/9506380];
G.~Bertone, D.~Hooper and J.~Silk,
  Phys.\ Rept.\  {\bf 405}, 279 (2005)
  [hep-ph/0404175].

\bibitem{Hinshaw:2012aka} 
  G.~Hinshaw {\it et al.} [WMAP Collaboration],
  Astrophys.\ J.\ Suppl.\  {\bf 208}, 19 (2013)
  [arXiv:1212.5226 [astro-ph.CO]].

\bibitem{Ade:2015xua} 
  P.~A.~R.~Ade {\it et al.} [Planck Collaboration],
  Astron.\ Astrophys.\  {\bf 594}, A13 (2016)
  [arXiv:1502.01589 [astro-ph.CO]].

\bibitem{kolbturner} E. W. Kolb and M. S. Turner, “The Early Universe", Addison-Wesley Pub. Company, 1989.

\bibitem{old_DM_Baryon_asy}
  S.~Nussinov,
  Phys.\ Lett.\  {\bf 165B}, 55 (1985);
K.~Griest and D.~Seckel,
  Nucl.\ Phys.\ B {\bf 283}, 681 (1987)
  Erratum: [Nucl.\ Phys.\ B {\bf 296}, 1034 (1988)];
R.~S.~Chivukula and T.~P.~Walker,
  Nucl.\ Phys.\ B {\bf 329}, 445 (1990);
S.~Dodelson, B.~R.~Greene and L.~M.~Widrow,
  Nucl.\ Phys.\ B {\bf 372}, 467 (1992);
S.~M.~Barr,
  Phys.\ Rev.\ D {\bf 44}, 3062 (1991);
D.~B.~Kaplan,
  Phys.\ Rev.\ Lett.\  {\bf 68}, 741 (1992);
H.~K.~Dreiner and G.~G.~Ross,
  Nucl.\ Phys.\ B {\bf 410}, 188 (1993)
  [hep-ph/9207221]; 
T.~Inui, T.~Ichihara, Y.~Mimura and N.~Sakai,
  Phys.\ Lett.\ B {\bf 325}, 392 (1994)
  [hep-ph/9310268];
S.~D.~Thomas,
  Phys.\ Lett.\ B {\bf 356}, 256 (1995)
  [hep-ph/9506274].


\bibitem{Asydm_models1}

R.~Kitano and I.~Low,
  Phys.\ Rev.\ D {\bf 71}, 023510 (2005)
  [hep-ph/0411133];  
K.~Agashe and G.~Servant,
  JCAP {\bf 0502}, 002 (2005)
  [hep-ph/0411254];
N.~Cosme, L.~Lopez Honorez and M.~H.~G.~Tytgat,
  Phys.\ Rev.\ D {\bf 72}, 043505 (2005)
  [hep-ph/0506320]; 
G.~R.~Farrar and G.~Zaharijas,
  Phys.\ Rev.\ Lett.\  {\bf 96}, 041302 (2006)
  [hep-ph/0510079];
R.~Kitano, H.~Murayama and M.~Ratz,
  Phys.\ Lett.\ B {\bf 669}, 145 (2008)
  [arXiv:0807.4313 [hep-ph]];
E.~Nardi, F.~Sannino and A.~Strumia,
  JCAP {\bf 0901}, 043 (2009)
  [arXiv:0811.4153 [hep-ph]];
H.~An, S.~L.~Chen, R.~N.~Mohapatra and Y.~Zhang,
  JHEP {\bf 1003}, 124 (2010)
  [arXiv:0911.4463 [hep-ph]];
T.~Cohen and K.~M.~Zurek,
  Phys.\ Rev.\ Lett.\  {\bf 104}, 101301 (2010)
  [arXiv:0909.2035 [hep-ph]];
J.~Shelton and K.~M.~Zurek,
  Phys.\ Rev.\ D {\bf 82}, 123512 (2010)
  [arXiv:1008.1997 [hep-ph]];
H.~Davoudiasl, D.~E.~Morrissey, K.~Sigurdson and S.~Tulin,
  Phys.\ Rev.\ Lett.\  {\bf 105}, 211304 (2010)
  [arXiv:1008.2399 [hep-ph]];
N.~Haba and S.~Matsumoto,
  Prog.\ Theor.\ Phys.\  {\bf 125}, 1311 (2011)
  [arXiv:1008.2487 [hep-ph]];
M.~R.~Buckley and L.~Randall,
  JHEP {\bf 1109}, 009 (2011)
  [arXiv:1009.0270 [hep-ph]]; 
P.~H.~Gu, M.~Lindner, U.~Sarkar and X.~Zhang,
  Phys.\ Rev.\ D {\bf 83}, 055008 (2011)
  [arXiv:1009.2690 [hep-ph]]; 
M.~Blennow, B.~Dasgupta, E.~Fernandez-Martinez and N.~Rius,
  JHEP {\bf 1103}, 014 (2011)
  [arXiv:1009.3159 [hep-ph]]; 
J.~McDonald,
  Phys.\ Rev.\ D {\bf 83}, 083509 (2011)
  [arXiv:1009.3227 [hep-ph]]; 
L.~J.~Hall, J.~March-Russell and S.~M.~West,
  arXiv:1010.0245 [hep-ph];
J.~J.~Heckman and S.~J.~Rey,
  JHEP {\bf 1106}, 120 (2011)
  [arXiv:1102.5346 [hep-th]];
M.~T.~Frandsen, S.~Sarkar and K.~Schmidt-Hoberg,
  Phys.\ Rev.\ D {\bf 84}, 051703 (2011)
  [arXiv:1103.4350 [hep-ph]];
S.~Tulin, H.~B.~Yu and K.~M.~Zurek,
  JCAP {\bf 1205}, 013 (2012)
  [arXiv:1202.0283 [hep-ph]].
  
\bibitem{Asydm_models2}
K.~Kohri, A.~Mazumdar and N.~Sahu,
  Phys.\ Rev.\ D {\bf 80}, 103504 (2009)
  [arXiv:0905.1625 [hep-ph]]; 
K.~Kohri, A.~Mazumdar, N.~Sahu and P.~Stephens,
  Phys.\ Rev.\ D {\bf 80}, 061302 (2009)
  [arXiv:0907.0622 [hep-ph]];
K.~Kohri and N.~Sahu,
  Phys.\ Rev.\ D {\bf 88}, 103001 (2013)
  [arXiv:1306.5629 [hep-ph]];
  M.~Ibe, S.~Matsumoto and T.~T.~Yanagida,
  Phys.\ Lett.\ B {\bf 708}, 112 (2012)
  [arXiv:1110.5452 [hep-ph]];
  M.~L.~Graesser, I.~M.~Shoemaker and L.~Vecchi,
  JHEP {\bf 1110}, 110 (2011)
  [arXiv:1103.2771 [hep-ph]];
  D.~Hooper, J.~March-Russell and S.~M.~West,
  Phys.\ Lett.\ B {\bf 605}, 228 (2005)
  [hep-ph/0410114];
  H.~Iminniyaz, M.~Drees and X.~Chen,
  JCAP {\bf 1107}, 003 (2011)
  [arXiv:1104.5548 [hep-ph]];
N.~Haba, S.~Matsumoto and R.~Sato,
  Phys.\ Rev.\ D {\bf 84}, 055016 (2011)
  [arXiv:1101.5679 [hep-ph]];
Z.~Kang, J.~Li, T.~Li, T.~Liu and J.~M.~Yang,
  Eur.\ Phys.\ J.\ C {\bf 76}, no. 5, 270 (2016)
  [arXiv:1102.5644 [hep-ph]];
K.~Blum, A.~Efrati, Y.~Grossman, Y.~Nir and A.~Riotto,
  Phys.\ Rev.\ Lett.\  {\bf 109}, 051302 (2012)
  [arXiv:1201.2699 [hep-ph]];
M.~Fujii and T.~Yanagida,
  Phys.\ Lett.\ B {\bf 542}, 80 (2002)
  [hep-ph/0206066];
T.~Banks, S.~Echols and J.~L.~Jones,
  JHEP {\bf 0611}, 046 (2006)
  [hep-ph/0608104];
T.~R.~Dulaney, P.~Fileviez Perez and M.~B.~Wise,
  Phys.\ Rev.\ D {\bf 83}, 023520 (2011)
  [arXiv:1005.0617 [hep-ph]];
T.~Cohen, D.~J.~Phalen, A.~Pierce and K.~M.~Zurek,
  Phys.\ Rev.\ D {\bf 82}, 056001 (2010)
  [arXiv:1005.1655 [hep-ph]];
B.~Dutta and J.~Kumar,
  Phys.\ Lett.\ B {\bf 699}, 364 (2011)
  [arXiv:1012.1341 [hep-ph]];
A.~Falkowski, J.~T.~Ruderman and T.~Volansky,
  JHEP {\bf 1105}, 106 (2011)
  [arXiv:1101.4936 [hep-ph]];
J.~March-Russell and M.~McCullough,
  JCAP {\bf 1203}, 019 (2012)
  [arXiv:1106.4319 [hep-ph]];
M.~L.~Graesser, I.~M.~Shoemaker and L.~Vecchi,
  arXiv:1107.2666 [hep-ph]; 
K.~Kamada and M.~Yamaguchi,
  Phys.\ Rev.\ D {\bf 85}, 103530 (2012)
  [arXiv:1201.2636 [hep-ph]]; 
D.~G.~E.~Walker,
  arXiv:1202.2348 [hep-ph]; 
B.~Feldstein and A.~L.~Fitzpatrick,
  JCAP {\bf 1009}, 005 (2010)
  [arXiv:1003.5662 [hep-ph]];
J.~March-Russell, J.~Unwin and S.~M.~West,
  JHEP {\bf 1208}, 029 (2012)
  [arXiv:1203.4854 [hep-ph]];
Y.~Cai, M.~A.~Luty and D.~E.~Kaplan,
  arXiv:0909.5499 [hep-ph];
H.~An, S.~L.~Chen, R.~N.~Mohapatra, S.~Nussinov and Y.~Zhang,
  Phys.\ Rev.\ D {\bf 82}, 023533 (2010)
  [arXiv:1004.3296 [hep-ph]];
C.~Kouvaris and P.~Tinyakov,
  Phys.\ Rev.\ Lett.\  {\bf 107}, 091301 (2011)
  [arXiv:1104.0382 [astro-ph.CO]];
M.~R.~Buckley,
  Phys.\ Rev.\ D {\bf 84}, 043510 (2011)
  [arXiv:1104.1429 [hep-ph]];
S.~Chang and L.~Goodenough,
  Phys.\ Rev.\ D {\bf 84}, 023524 (2011)
  [arXiv:1105.3976 [hep-ph]];
S.~Profumo and L.~Ubaldi,
  JCAP {\bf 1108}, 020 (2011)
  [arXiv:1106.4568 [hep-ph]];
H.~Davoudiasl, D.~E.~Morrissey, K.~Sigurdson and S.~Tulin,
  Phys.\ Rev.\ D {\bf 84}, 096008 (2011)
  [arXiv:1106.4320 [hep-ph]];
I.~Masina and F.~Sannino,
  JCAP {\bf 1109}, 021 (2011)
  [arXiv:1106.3353 [hep-ph]];
T.~Lin, H.~B.~Yu and K.~M.~Zurek,
  Phys.\ Rev.\ D {\bf 85}, 063503 (2012)
  [arXiv:1111.0293 [hep-ph]];
M.~R.~Buckley and S.~Profumo,
  Phys.\ Rev.\ Lett.\  {\bf 108}, 011301 (2012)
  [arXiv:1109.2164 [hep-ph]];
H.~Davoudiasl and R.~N.~Mohapatra,
  New J.\ Phys.\  {\bf 14}, 095011 (2012)
  [arXiv:1203.1247 [hep-ph]];
N.~Okada and O.~Seto, %
``Originally Asymmetric Dark Matter,''
 Phys.\ Rev.\ D {\bf 86}, 063525 (2012)
[arXiv:1205.2844 [hep-ph]];
  T.~Hugle, M.~Platscher and K.~Schmitz,
  arXiv:1804.09660 [hep-ph].
 
\bibitem{Asydm_models3}
C.~Arina and N.~Sahu,
  Nucl.\ Phys.\ B {\bf 854}, 666 (2012)
  [arXiv:1108.3967 [hep-ph]];
C.~Arina, J.~O.~Gong and N.~Sahu,
  Nucl.\ Phys.\ B {\bf 865}, 430 (2012)
  [arXiv:1206.0009 [hep-ph]];
C.~Arina, R.~N.~Mohapatra and N.~Sahu,
  Phys.\ Lett.\ B {\bf 720}, 130 (2013)
  [arXiv:1211.0435 [hep-ph]]. 

\bibitem{Asydm_review} See for a review:
K.~Petraki and R.~R.~Volkas,
  Int.\ J.\ Mod.\ Phys.\ A {\bf 28}, 1330028 (2013)
  [arXiv:1305.4939 [hep-ph]];
K.~M.~Zurek,
  Phys.\ Rept.\  {\bf 537}, 91 (2014)
  [arXiv:1308.0338 [hep-ph]].

\bibitem{Patrignani:2016xqp} 
  C.~Patrignani {\it et al.} [Particle Data Group],
  Chin.\ Phys.\ C {\bf 40}, no. 10, 100001 (2016).

\bibitem{Sakharov:1967dj} 
  A.~D.~Sakharov,
  Pisma Zh.\ Eksp.\ Teor.\ Fiz.\  {\bf 5}, 32 (1967)
  [JETP Lett.\  {\bf 5}, 24 (1967)]
  [Sov.\ Phys.\ Usp.\  {\bf 34}, 392 (1991)]
  [Usp.\ Fiz.\ Nauk {\bf 161}, 61 (1991)].

\bibitem{fukugita&Yanagida_1986} M.Fukugita and T. Yanagida, Phys.\ Lett.\ B {\bf 174}, 45 (1986).  

\bibitem{Kaplan:2009ag} 
  D.~E.~Kaplan, M.~A.~Luty and K.~M.~Zurek,
  Phys.\ Rev.\ D {\bf 79}, 115016 (2009)
  [arXiv:0901.4117 [hep-ph]].

\bibitem{Feng:2012jn} 
  W.~Z.~Feng, P.~Nath and G.~Peim,
  Phys.\ Rev.\ D {\bf 85}, 115016 (2012)
  [arXiv:1204.5752 [hep-ph]].
  
\bibitem{Ibe:2011hq}
  M.~Ibe, S.~Matsumoto and T.~T.~Yanagida,
  Phys.\ Lett.\ B {\bf 708} (2012) 112
  [arXiv:1110.5452 [hep-ph]].

\bibitem{type1_seesaw}
P. Minkowski, Phys. Lett. {\bf B 67}, 421 (1977);
M.~Gell-Mann, P.~Ramond and R.~Slansky in {\it Supergravity} (P.~van Niewenhuizen and D.~Freedman, eds),
(Amsterdam), North Holland, 1979; 
T.~Yanagida in {\it Workshop on Unified Theory and Baryon number in the Universe} (O. Sawada
and A.~Sugamoto, eds), (Japan), KEK 1979; R.N.~Mohapatra and
G.~Senjanovic, Phys.\ Rev.\ Lett. {\bf 44}, 912 (1980).

\bibitem{Geng:1988pr} 
  C.~Q.~Geng and R.~E.~Marshak,
  Phys.\ Rev.\ D {\bf 39}, 693 (1989).
X.~G.~He, G.~C.~Joshi and R.~R.~Volkas,
  Phys.\ Rev.\ D {\bf 41}, 278 (1990).
K.~Kohri and N.~Sahu,
  Phys.\ Rev.\ D {\bf 88}, 103001 (2013)
  [arXiv:1306.5629 [hep-ph]].

\bibitem{Cai:2017jrq} 
  Y.~Cai, J.~Herrero-García, M.~A.~Schmidt, A.~Vicente and R.~R.~Volkas,
  Front.\ in Phys.\  {\bf 5}, 63 (2017)
  [arXiv:1706.08524 [hep-ph]].

\bibitem{buchmuller&plumacher} W. Buchmuller and M. Plumacher, Phys. Rept. 320, 329 (1999).

\bibitem{Buchmuller:2004nz} 
  W.~Buchmuller, P.~Di Bari and M.~Plumacher,
  Annals Phys.\  {\bf 315}, 305 (2005)
  [hep-ph/0401240].
  
\bibitem{Giudice:2003jh} 
  G.~F.~Giudice, A.~Notari, M.~Raidal, A.~Riotto and A.~Strumia,
  Nucl.\ Phys.\ B {\bf 685}, 89 (2004)
  [hep-ph/0310123].  
  
\bibitem{Bernal:2016gfn} 
  N.~Bernal, C.~S.~Fong and N.~Fonseca,
  JCAP {\bf 1609}, no. 09, 005 (2016)
  [arXiv:1605.07188 [hep-ph]].

\bibitem{Harvey:1990qw} 
  J.~A.~Harvey and M.~S.~Turner,
  Phys.\ Rev.\ D {\bf 42}, 3344 (1990).


\bibitem{Khachatryan:2016whc} 
  V.~Khachatryan {\it et al.} [CMS Collaboration],
  JHEP {\bf 1702}, 135 (2017)
  [arXiv:1610.09218 [hep-ex]].


\bibitem{Aprile:2012nq} 
  E.~Aprile {\it et al.} [XENON100 Collaboration],
  Phys.\ Rev.\ Lett.\  {\bf 109}, 181301 (2012)
  [arXiv:1207.5988 [astro-ph.CO]].

\bibitem{Akerib:2016vxi} 
  D.~S.~Akerib {\it et al.} [LUX Collaboration],
  Phys.\ Rev.\ Lett.\  {\bf 118}, no. 2, 021303 (2017)
  [arXiv:1608.07648 [astro-ph.CO]].

\bibitem{Aprile:2015uzo} 
  E.~Aprile {\it et al.} [XENON Collaboration],
  JCAP {\bf 1604}, no. 04, 027 (2016)
  [arXiv:1512.07501 [physics.ins-det]].


\bibitem{Goodman:1984dc} 
  M.~W.~Goodman and E.~Witten,
  Phys.\ Rev.\ D {\bf 31}, 3059 (1985).

\bibitem{Ellis:2008hf} 
  J.~R.~Ellis, K.~A.~Olive and C.~Savage,
  Phys.\ Rev.\ D {\bf 77}, 065026 (2008)
  [arXiv:0801.3656 [hep-ph]].

\bibitem{Akrami:2010dn} 
  Y.~Akrami, C.~Savage, P.~Scott, J.~Conrad and J.~Edsjo,
  JCAP {\bf 1104}, 012 (2011)
  [arXiv:1011.4318 [astro-ph.CO]].

\bibitem{NSNS} 
  S.~Bhattacharya, N.~Sahoo and N.~Sahu,
  Phys.\ Rev.\ D {\bf 93}, no. 11, 115040 (2016)
  [arXiv:1510.02760 [hep-ph]];

S.~Patra, S.~Rao, N.~Sahoo and N.~Sahu,
  Nucl.\ Phys.\ B {\bf 917}, 317 (2017)
  [arXiv:1607.04046 [hep-ph]];

S.~Bhattacharya, N.~Sahoo and N.~Sahu,
  Phys.\ Rev.\ D {\bf 96}, no. 3, 035010 (2017)
  [arXiv:1704.03417 [hep-ph]].

\bibitem{Ellis:2000ds} 
  J.~R.~Ellis, A.~Ferstl and K.~A.~Olive,
  Phys.\ Lett.\ B {\bf 481}, 304 (2000)
  [hep-ph/0001005].

\bibitem{cms_report_2018} CMS collaboration, CMS-PAS-HIG-17-031, March 2018.

\bibitem{Khachatryan:2016vau} 
  G.~Aad {\it et al.} [ATLAS and CMS Collaborations],
  JHEP {\bf 1608}, 045 (2016)
  [arXiv:1606.02266 [hep-ex]].

\bibitem{Angloher:2015ewa} 
  G.~Angloher {\it et al.} [CRESST Collaboration],
  Eur.\ Phys.\ J.\ C {\bf 76}, no. 1, 25 (2016)
  doi:10.1140/epjc/s10052-016-3877-3
  [arXiv:1509.01515 [astro-ph.CO]].

\end{thebibliography}
\end{document}